\documentstyle[pre%print
,aps]{revtex}
\def\grl{{Geophys.~Res.~Lett.}}
\def\apjs{{\apj\ Suppl.}}			
\def\aap{{Astronomy and Astrophys.}}
\def\asr{{Adv. Space Res. }}	

\def\pr{{Phys.~Rev.}}	
\def\prl{{Phys.~Rev.~Lett.}}	
\def\jgr{{J.~Geophys.~Res.}}
\def\physrep{{Phys.~Rep.}}
\def\mnras{{MNRAS}}		
\def\onehalf{\slantfrac{1}{2}}	
\def\ga{\gtrsim}
\def\la{\lesssim}
\def\etal{{\it et al.,\ }}
\def\eg{{ e.g.,\ }}
\def\ie{{ i.e.,\ }}
\def\mathbf{\bf}
\def\mathrm{\rm}
\def\mathcal{\cal}

\twocolumn
\begin{document}
\draft

\title{
 Ion leakage from  quasiparallel collisionless shocks: \protect\\
implications for injection and shock dissipation}
\author{M.A. Malkov}
\address{Max-Planck Institut f\"ur Kernphysik, D-69029, Heidelberg, Germany}
\date{\today}
%% Email: Malkov@boris.mpi-hd.mpg.de
 
\maketitle
\begin{abstract}
 A simplified model of particle transport at a quasiparallel
 one-dimensional collisionless shock is suggested.  In this model the
 MHD-turbulence behind the shock is dominated by a circularly
 polarized, large amplitude Alfv\'en wave originated upstream from the
 turbulence excited by particles leaking from the downstream medium.
 It is argued that such a wave having significantly increased its
 magnetic field during the transmission through the shock interface
 can effectively trap thermal ions, regulating their leakage upstream.
 Together with a background turbulence this wave also plays a
 fundamental role in thermalization of the incoming ion flow.  The
 spectrum of leaking particles and the amplitude of the wave excited
 by these particles are selfconsistently calculated.  The injection
 rate into the first order Fermi acceleration based on this leakage
 mechanism is obtained and compared with computer simulations. The
 related problem of shock energy distribution between  thermal and
 nonthermal components of the shocked plasma is discussed.  The
 chemical composition of the leaking particles is studied.
\end{abstract}

\section{Introduction}\label{intr}
The problem of energy dissipation in collisionless shocks in plasmas is old and 
exceedingly difficult \cite{park61,sag66,ks67,cor70,lee82,keh85,pap85,sch,quest}.  
Moreover, there exist persuasive theoretical arguments \cite{als77,dv81} 
corroborated by numerous simulations (\eg \cite{je}) that this problem cannot be 
solved by considering exclusively the thermalization of the bulk plasma flow 
when the latter passes through the shock.  A significant part if not almost all 
the energy of a strong large shock may be channeled into a small minority of 
particles accelerated through multiple crossing of its front.  This acceleration 
mechanism is known as the first order Fermi or diffusive shock acceleration.  
However, this does not circumvent the problem of collisionless thermalization.  
The reason is that a small fraction of thermal ions that leak or reflect from 
the shock play a crucial role in the collisionless energy exchange between the 
bulk upstream motion and thermal and/or nonthermal (accelerated) components of 
the downstream plasma.  These ions generate waves in the foreshock region whose 
growthrate and amplitude are directly related to their density.  
Yet they provide a seed, or injection population for the further acceleration.  
These two aspects of the shock dissipation are clearly interrelated.  By wave 
excitation these particles create a scattering environment allowing them to 
cross the shock repeatedly which is necessary for the Fermi mechanism to work.

One of the most important parameters of collisionless shocks is the
angle $\theta_{\rm nB}$ between an ambient magnetic field and the
shock normal.  While so called perpendicular shocks ($\theta_{\rm nB}
\simeq \pi/2$) should clearly have a distinct shock transition because
the hot downstream plasma cannot penetrate upstream for more than one
ion gyroradius, their parallel counterparts ($\theta_{\rm nB} \ll 1$)
are not so suitable for confinement of the heated downstream plasma
since it may penetrate far upstream moving along the field lines.  We
will confine our consideration below to this latter category of
collisionless shocks.  In general, shocks with $\theta_{\rm nB} <
\pi/4$ are somewhat superficially referred to as quasiparallel whereas
the rest ($\pi/4 < \theta_{\rm nB} < \pi/2 $) as quasi-perpendicular.

It is important to realize that the distribution function of the
backstreaming particles cannot be inferred solely from the macroscopic
parameters of the downstream plasma even if the thermalization
mechanism is properly understood.  One obvious reason for this is the
following.  Among the backstreaming particles one can find not only
those which simply arose from the randomization of the upstream flow
at the shock (as backscattered from the downstream medium or reflected
from the shock interface) but also the particles of these two types
which crossed the shock more than once and therefore have gained some
energy \cite{mv95}.

The above arguments suggest that the injection problem can be divided
into the following two tasks: (i) given the shock conditions one
determines the distribution of particles originating from the upstream
flow after they crossed the shock for the first time.  Subsequently,
one identifies those particles which are capable of crossing the shock
also in reverse direction (first generation of injected particles).
(ii) One follows the (stochastic) trajectories of these particles when
they multiply recross the shock until they are swept downstream or
have achieved energies acceptable for the standard description of
diffusive shock acceleration (see \eg Ref.  \cite{bell78} or
\cite{dru83} for a review).

The first task belongs to collisionless shock physics \cite{keh85,pap85} and can 
at least formally be treated independently of the diffusive shock acceleration 
process.  The second constitutes the injection problem itself as a part of 
diffusive shock acceleration theory and can be formulated in more detail as 
follows.  Suppose the task (i) is solved.  Then, given the distribution of 
thermal particles that are able to penetrate into the upstream region, one 
calculates the high energy asymptotics of their distribution.  This provides the 
coefficient in the power-law solution of the standard acceleration theory and 
thus the injection rate.  It cannot be obtained within the standard theory since 
the latter is unable to describe low energy particles with anisotropic pitch 
angle distribution.

The solution of the injection problem (ii) as formulated above has been obtained 
analytically in Ref.  \cite{mv95}.  The high energy asymptotics of this solution 
indeed matches the power-law of the standard theory.  At the lower energy end it 
smoothly joins the downstream thermal distribution.  This thermal distribution 
that determines the flux of leaking particles has been simply assumed to be 
created by a unspecified shock randomization process, in other words due to the 
action of a ``thermostat''.  In this paper we suggest a simplified thermostat 
model.  In the next subsection we put it in the general context of collisionless 
shocks.
\subsection{Thermostat model}\label{intr1} 
Some of the thermostat properties are readily known from ordinary 
Rankine-Hugoniot (RH) jump conditions that constitute the conservation of mass, 
momentum and energy fluxes across the shock \cite{ll:hd}.  In the simplest case 
when both the magnetic field and the flow velocity are perpendicular to the 
shock front the RH conditions are the same as in ordinary gasdynamics
\begin{eqnarray}
	\rho_1 u_1 & = & \rho_2 u_2
	\label{c:e}  \\
	\rho_1 u_1^2 +P_1 & =&\rho_2 u_2^2+P_2  
	\label{ber}  \\
	\frac{1}{2}\rho_1 u_1^3 +\frac{\gamma}{\gamma-1}P_1 u_1& =
	 & \frac{1}{2}\rho_2 u_2^3 +\frac{\gamma}{\gamma-1}P_2 u_2
	\label{en}
\end{eqnarray}
Here the index 1 (2) refers to the upstream (downstream) medium, $\rho\simeq 
m_{\rm p}n, \, u, $ and $P $ are the mass density, velocity and the gas kinetic 
pressure, respectively, $m_{\rm p}$ is the proton mass.  The adiabatic index 
$\gamma$ can be set to $\gamma=5/3$.  The pressure and energy of the MHD waves 
that are implied to participate in the shock thermalization process as a 
substitute for the binary collisions are neglected here for the following 
reasons.  First, if the initial state (1) refers to the far upstream region in 
which the backscattering particles do not penetrate and the waves are thus not 
present there, the left hand sides of the above equations are written correctly.  
If the final state (2) is also taken so far downstream where the thermalization 
process is completed and the energy of the waves excited at the shock interface 
is damped, the same is true for the right hand sides.  Even if the final and 
initial states are taken substantially closer to the shock interface these 
equations correctly describe the jump relations in the case of a very strong 
shock.  Indeed, the waves that are excited in a highly supersonic and 
superalfvenic upstream flow via cyclotron resonance with the beam of 
backstreaming particles, are the transverse ($\delta {\mathbf{B}} \perp 
{\mathbf{B}}_0$) MHD waves and their amplitude $\delta B$ cannot exceed the 
unperturbed field $B_0$ since otherwise the beam would be trapped in this wave 
and carried back downstream.  Therefore, considering strong shocks with $M_{\rm 
A}^2 \equiv u_1^2/v_{\rm A}^2\equiv 4 \pi \rho_1 u_1^2/B_0^2 \gg 1$ we may 
neglect the contribution of the wave energy upstream.  Downstream, it is 
increased by a factor $\sim (u_1/u_2)^2$ due to the shock compression of the 
perpendicular component of the wave magnetic field but the wave energy may still 
be neglected compared to the thermal and the bulk motion energy in strong shocks 
with $M_{\rm A} \gg 1$.  For whatever reasons the RH conditions have been 
observationally proven to be quite reliable in collisionless shocks 
\cite{kenetal84,rus:far}.

In the case of high Mach number shocks $M_{\rm S}^2 \equiv
u_1^{2}\rho_1^{2}/\gamma P_1 \gg 1$ the RH conditions simply yield
(for $\gamma=5/3$): $u_1/u_2=4$ and the downstream temperature $
T_2=3m_{\rm p}u_2^2$.  If one is assuming that all the downstream
particles that have velocities $v_{\mathrm{z}} < - u_2$ (where
$v_{\mathrm{z}}$ denotes the normal to the shock front velocity
component, and the shock is moving in negative \( z- \) direction) can
cross the shock, this information about the thermostat suffices to
obtain an injection solution.  On the other hand, such an assumption
can only give an upper bound on the injection efficiency and it highly
overestimates the latter. In principle, many particles in such a
high temperature downstream distribution have negative velocities in
the shock frame of reference and, therefore, can potentially escape
upstream.  But this would be a purely kinematic picture.  In reality
too intensive escape would result in a fast excitation of waves
scattering the beam back downstream over a short distance.  This
process can be understood within the framework of the quasilinear
theory of the cyclotron instability developed for homogeneous plasma
in Ref.  \cite{rss} (see also \cite{mv95} and Sec.  \ref{sc} below).

Consider the phase space of such a beam emanating from the shock
upstream on the plane $(v_{\parallel},v_{\perp})$ in the shock frame
of reference Fig.  \ref{b:ps}, where $(v_{\parallel},v_{\perp})$ are
parallel and perpendicular to the shock normal (and to the unperturbed
magnetic field) components of particle velocity.  The arrangement of
the beam in the phase space corresponds qualitatively to its
parameters calculated later in this paper.  As soon as the beam
appears on the upstream side of the shock it starts to generate MHD
waves that move to the left in the local plasma frame at the Alfv\'en
speed $-C_{\rm A}$ but are in fact continuously convected back to the
shock with the main flow since $u_1 \gg C_{\rm A}$.  Their resonance
length is of the order of $u_1/\omega_{\rm ci}$ where $\omega_{\rm ci
}$ is the ion cyclotron frequency.  The beam itself is being scattered
in pitch angle by this selfgenerated waves around the center at
$u_1-C_{\rm A}$ so that it gradually diffuse into the $v_{\parallel} >
0 $ region which means that it returns downstream.  The relaxation
length of the beam, \ie the depth of its penetration upstream may be
estimated as $l_{\rm R} \sim u_1/\gamma$ where $\gamma$ is the growth
rate of the cyclotron instability upstream.  Estimating $\gamma$ from
quasilinear equations (\eg  Ref.  \cite{rss}) one may obtain
for~$l_{\rm R}$
$$l_{\rm R}\sim \frac{c}{\omega_{\rm pi}}\frac{n_0}{n_{\rm
b}}\frac{\Delta v_{\parallel}}{u_1}$$ where $\omega_{\rm pi}$ is the
ion plasma frequency, $n_{\rm b},\, n_0$ are the beam and the
background densities, respectively, and $\Delta v_{\parallel}$ is the
beam width in $v_{\parallel}$. The last factor in $l_{\rm R}$ is in
fact only slightly less than unity so that the upstream part of the
entire shock transition is roughly $n_0/n_{\rm b} \gg 1$ ion inertial
lengths, $c/\omega_{\rm pi}$.

In principle the turbulence generated by the beam and growing in the
downstream direction could gradually saturate at a distance $\ga
l_{\rm R}$ and the plasma flow beyond this distance could be declared
as a downstream state.  However, this is not what actually happens as
both the observations and numerical simulations of strong shocks
reveal (see \eg \cite{pap85}, \cite{quest}).  Namely, there exists a
relatively sharp shock transition inside of this structure where the
amplitude of magnetic pulsations increases over a distance $\sim
c/\omega_{\rm pi}$, the bulk flow slows down nearly to its downstream
value and particle orbits spread in velocity to an approximately
thermal width in a substantially increased magnetic field.

Existing 1D and 2D hybrid simulations (ions treated as particles and electrons 
as a fluid) of quasiparallel shocks \cite{quest,lk90,k-vo93,sch93,be95,skj} also 
provide some clue to how particles are confined on the downstream side of the 
shock.  Namely, despite the fact that there are indeed many particles just 
downstream of the shock front that have velocities \( v_\parallel < 0 \), only 
few of them recross the shock.  This suggests that the heated downstream plasma 
is directly locked by the waves which have been excited in the upstream region 
and transmitted then across the shock.  In part, they may be generated at the 
shock interface.  Because of a substantial downstream increase of the wave 
amplitude due to the shock compression, the perpendicular component of \( B- 
\)field may become large enough to trap the bulk of the particles and to sweep 
them downstream.  An additional factor that can also help to confine the shocked 
plasma is an electrostatic barrier that is also observed in simulations.  The 
analysis of the backscattered ions performed by Quest \cite{quest} shows that 
the upstream flux of these particles is significantly smaller than it would be 
if all downstream particles with negative velocities were to stream freely 
across the shock.

Another important conclusion that can be drawn from the 1D hybrid simulations 
(see \eg Ref.  \cite{quest}) is that the downstream wave turbulence is dominated 
by a mode excited by the backstreaming beam in the upstream region.  One can 
think of a circularly polarized quasi-monochromatic MHD wave convected 
downstream with the bulk plasma and extended there over a distance of the order 
of hundred ion inertial lengths.  The amplitude of this wave is quite large 
$\delta B/B_0 \sim 3-4$ (see also Ref.  \cite{be95}).  Traces of the 
quasi-regular particle motion in the wave are also seen in the simulations 
\cite{quest}.  Beyond this distance the wave is gradually damped and the 
particle distribution tends to a thermal one.  However, unlike the upstream part 
$l_{\rm R}$ of the total shock transition the distance where it happens is very 
difficult to assess analytically.  The reason is that the level of turbulence is 
very high and a weakly turbulent approach is probably impossible (see, however, 
Sec.  \ref{bt}).  This part of the shock transition should rather be regarded as 
a BGK type of wave \cite{bgk}.  Such an attempt is made in a companion paper 
\cite{m97c} where we consider the slowing down and the heating of the upstream 
plasma caused by interaction with this wave downstream.

The present paper suggests a model of ion leakage allowing the determination of 
the thermostat production rate that was only parameterized in the injection 
theory \cite{mv95}.  The starting point of the model is the particle dynamics in 
the Alfv\'en wave behind the shock \cite{f1}.  The particle phase space is 
divided into two parts.  One part contains the ``trapped'' particles which are 
convected downstream with the wave.  The second part contains ``untrapped'' or 
adiabatically with the wave interacting particles, and particles trapped in 
the nonlinear Doppler resonance.  These particles can escape upstream when their 
averaged velocity with respect to the wave is high enough.

It is not assumed, however, that particle motion in this model is purely 
regular.  Instead, a downstream turbulence that should exist together with the 
monochromatic wave allows particles to cross the boundary between trapped and 
untrapped regions and also between the ``lock'' and ``escape'' states of 
particles in the phase space.  Although this motion across the invariant 
manifolds of the regular dynamics is assumed to be relatively slow it should 
produce the necessary entropy by virtue of the standard arguments 
\cite{prig62,sg69}: even a very weak perturbation can quickly randomize the 
particle motion while imposed onto the regular motion in the large amplitude 
wave since the main job is done by this wave via fast phase mixing.  From this 
reasoning we obtain the fraction of the downstream particles that escape 
upstream from such a trap as a function of their energy.  This allows us to 
calculate the unknown ``thermostat'' distribution function in terms of the wave 
amplitude.

The final step of this scheme is the determination of the wave amplitude from 
the beam density.  This provides, in fact, an equation for the wave amplitude 
since the beam density depends on this as well.  Upon solution of this 
equation one obtains both the beam density and the wave amplitude as functions 
of shock parameters.

In the next section we briefly review the particle dynamics in a monochromatic 
circularly polarized wave, introduce suitable variables, and consider the 
special case of a very strong wave ($\delta B/B_0 \gg 1$).  In Sec.  \ref{bt} 
the role of the background turbulence is discussed and adiabatically leaking 
particles are identified.  In Sec.  \ref{inv:mes} the probability of their 
escape is calculated as a function of energy and wave amplitude.  In Sec.  
\ref{re} the resonant escape is considered.  Sec.  \ref{a:z} deals with the 
dependence of escape fluxes upon the mass to charge ratio.  In Sec.  \ref{sc} we 
calculate the wave amplitude upstream.  This enables us to obtain the particle 
flux without parameterization, simply as a function of the Mach number.  After a 
brief summary of the considered leakage mechanism in Sec.  \ref{ns} we calculate 
in Sec.  \ref{comp} the spectrum injected into the first order Fermi 
acceleration and compare it with the hybrid simulations.  We conclude with a 
discussion of alternative mechanisms and possible applications to the problem of 
energy partition of collisionless shocks between thermal and nonthermal 
particles.  
%%***********************************************
\section{Particle dynamics in the wave}\label{pd}
%%***********************************************
The equations of motion in the case considered here are fully
equivalent to the equations of particle motion in a whistler wave.
These equations have been extensively studied in the past.  It is well
known \cite{lut:sud} that they are completely integrable.  However,
unlike the situation with the whistler wave where one usually assumes
$\delta B/B_0 \ll 1$, we must concentrate on the opposite case as it
was described in the preceding section. We therefore present here a
technically different description of the particle dynamics which is
suitable to our purposes.

Let us assume that the wave propagates in $z$- direction, i.e.  normal
to the shock and $B_0=B_{\rm z}=const$.  We represent the total
magnetic field in the reference frame moving with the wave in the
following form 
\begin{equation} 
{\mathbf B} = B_{\mathrm{z}} {\mathbf
e}_{\mathrm z} + B_{\perp} \left(-{\mathbf e}_{\mathrm{x}} \cos k_0 z
+ {\mathbf e}_{\mathrm{y}}\sin k_0 z\right) 
\label{mf} 
\end{equation}
where $\{\mathbf{e}_{\mathrm{i}}\} $ denotes the standard basis in
coordinate space and $ k_0 $ is the wave number.  The electric field
vanishes in the wave frame of reference and the equations of motion
read
\begin{eqnarray}
	\frac{d\mathbf{ v}}{d t} & = & \frac{e}{m_{\rm p} c}\mathbf{v} \times 
	\mathbf{B} \label{eqm1}  \\
\frac{d z}{d t} &  = &v_{\mathrm{z}}
	\label{v_z}
\end{eqnarray}
   where the particle velocity ${\mathbf v} = v_{\mathrm{x}}
  {\mathbf e}_{\rm x} + v_{\rm y}{\mathbf e}_{\rm y}+ v_{\rm z}{\mathbf
  e}_{\rm z} $.  Interested in the case
  \begin{equation}
	\varepsilon   \equiv \frac{B_{\rm z}}{B_{\perp}} \ll 1,
	\label{eps}
  \end{equation} 
  we introduce the cyclotron frequency
  \begin{equation}
	\omega_{\perp} = \frac{e B_{\perp}}{m_{\rm p} c}
	\label{om_h}
   \end{equation}
and rescale the variables in Eqs.\  (\ref{eqm1},\ref{v_z}) as follows
   \begin{equation}
     	k_0 z \rightarrow z, \quad \omega_{\perp} t \rightarrow t, \quad 
     	\frac{k_0 {\mathbf v}}{\omega_{\perp}} \rightarrow {\mathbf v}.
     	\label{dimlvar}
     \end{equation}
Using these new variables Eq.(\ref{eqm1}) rewrites as:
\begin{equation}
	 \frac{d {\bf v}}{d t}={\bf v}\times \left(-{\bf e}_{\rm x} \cos z + 
{\bf e}_{\rm y} \sin z + \varepsilon {\bf e}_{\rm z}\right) 
	\label{dimleq}
\end{equation} 
and Eq. (\ref{v_z}) remains unchanged.
It is convenient to make  a further transformation in these equations
$\left(v_{\rm x},v_{\rm y},v_{\rm z},z \right) \rightarrow \left(\lambda, 
q, v_{\rm z},z \right)$, in which
\begin{equation}
	v_{\rm x}+iv_{\rm y} =  -\sqrt{\lambda^2 - 2 \varepsilon v_{\rm z}} 
\exp \left[i(q - z)\right] + 
\exp (-iz) 
	\label{vx:vy}
\end{equation}
 It is easy to see that $\lambda $ is conserved,  
$ {d \lambda}/{d t} = 0 $, 
and we arrive at the following one dimensional (i.e.  integrable) dynamical 
system in the variables $(q,v_{\rm z})$
\begin{eqnarray}
	\frac{d q}{d t} & = &  v_{\rm z}+\frac{\varepsilon \cos q}
{\sqrt{\lambda^2 - 2 \varepsilon v_{\rm z}}}- \varepsilon \nonumber \\
	\frac{d v_{\rm z}}{d t} & = & -\sqrt{\lambda^2 - 2 \varepsilon 
	v_{\rm z}} \sin q
	\label{dq:dvz}
	\end{eqnarray}
The integral  $\lambda^2 $ can be written in the old variables as
$$	\lambda ^2 = v_{\rm x}^2 + v_{\rm y}^2 - 2 v_{\rm x} \cos z + 2 
	v_{\rm y}\sin z + 2 \varepsilon v_{\rm z} +1 $$
Besides $\lambda $ there is another obvious integral, the energy
$	v^2 = v_{\rm x}^2+v_{\rm y}^2+v_{\rm z}^2 $.
Note that $\lambda^2 $ in Eq.(\ref{dq:dvz}) can be negative in certain 
parts of the phase space; positive definite is only the quantity $\lambda^2 -2 
\varepsilon v_{\rm z}$.  However, being interested in the case $\varepsilon 
\ll 1 $, we start our consideration of the system (\ref{dq:dvz}) from the 
simplest situation where $\lambda^2 \gg \left|\varepsilon v_{\rm 
z}\right|$.  As we will see the escaping particles interact with the wave 
adiabatically (\( v > 1 \)) in this case.  We shall return to the case of 
small and negative $\lambda^2 \la \varepsilon$ (resonant escape) in Sec.  
\ref{re}.
%%##############
\subsection{Adiabatic wave-particle interaction}\label{ad:wpi}
Introducing the new variable 
\begin{equation}
	\eta = v_{\rm z}+\frac{\varepsilon}{\lambda} \cos q -\varepsilon
	\label{eta}
\end{equation}
and retaining only the terms of zeroth and first order in $\varepsilon $, 
the system (\ref{dq:dvz}) describes a simple pendulum
\begin{equation}
	\frac{d q}{d t}  =  \eta;\quad
	\frac{d \eta}{d t} = -\lambda \sin q
\label{dq:dt}
\end{equation}
with the Hamiltonian 
\begin{equation}
	H = 2 \lambda \sin ^2 (\frac{q}{2}) + \frac{1}{2} \eta ^2.
	\label{Ham}
\end{equation}
that is connected with the integrals $v $ and 
$\lambda $ by means of the following relation
$	v^2 = 2 H + (\lambda -1)^2 $. 
It is  convenient to introduce the standard action-angle variables in 
the system (\ref{dq:dt}-\ref{Ham}), using the truncated action 
\begin{equation}
	S = \int \eta d q
	\label{S}
\end{equation}
as a generating function. The function 
$	k^2 = 2 {\lambda}/{H} = {4 \lambda}/[v^2 - (\lambda - 1)^2] $
divides particle phase space into two parts that will be superficially 
referred to as the region of trapped $(k > 1) $ and untrapped $(k < 1) $ 
particles.  It should be pointed out that the particles with $k > 1 $ are 
not really trapped in the usual sense because they can become untrapped $(k 
< 1) $ without changing their energy $v^2 $.  We shall return to this point 
below.  Here we only note that the analogy to \eg the particle dynamics in 
a monochromatic Langmuir wave \cite{on} is incomplete in this respect.  In 
many other respects ``trapped'' particles behave as such, in particular 
their averaged velocity $\bar \eta $ is zero.  According to Eq. (\ref{eta}) 
this means that $\bar v_{\rm z} \sim \varepsilon $.

For the untrapped particles, using equations (\ref{Ham},\ref{S}), we obtain 
\begin{equation}
	S = 4 \frac{\sqrt{\lambda}}{k} E\left(\frac{q}{2},k\right) 
	\label{act:untr}
\end{equation}
where $ E $ is the incomplete elliptic integral of the second kind.  
It is convenient to define an action $J$ as 
\begin{equation}
	J=\frac{1}{2 \pi}\, {\rm sign } (\eta )\int_{0}^{2 \pi} \eta d q
	\label{I:trd}
\end{equation}
so that we finally obtain
  \begin{equation}
  	J = \frac{4 \sqrt{\lambda}}{\pi k} {\rm sign } (\eta) {\bf E}(k),
  	\label{I:tr}
  \end{equation}
where ${\bf E } $ denotes the complete elliptic integral of the second 
kind.  Thus, the untrapped particles occupy the regions $\left| J \right| 
\ge 8\sqrt{\lambda}/\pi \equiv J_{\rm S}$, and far from the separatrix 
($k=1$) where $k \rightarrow 0 $ one simply has $J \rightarrow \eta $.  The 
angle variable $\psi $ conjugate to $J $ is
\begin{equation}
	\psi = \frac{\partial S}{\partial J} = \pi 
\frac{F\left(\frac{q}{2},k\right)}{{\bf K}\left(k\right)}
	\label{psi}
\end{equation}
where $F $ and ${\bf K} $ are the incomplete and complete elliptic 
integrals of the first kind. Note that \( \psi \rightarrow 
q \) as \( k \rightarrow 0 \).

We will not use the corresponding action-angle variables for trapped 
particles.
%%****************************************************** 
\section{The background turbulence and escape upstream}\label{bt}
The simple particle dynamics in a monochromatic wave considered in the 
preceding section implies that the amplitude of this wave is constant in 
space and time.  This is certainly not the case for the wave associated 
with a shock.  As we mentioned in the Introduction the wave is extended 
over a finite distance on the downstream side of the shock (we assume 
between $z = 0 $ and $z = L $, and $k_0 L \gg 1$) and decays at larger $z$.  
Thus,  particle interaction with this wave occurs in the following way.  
First, particles that cross $z = 0 $ plane from $z < 0 $ become 
trapped (at least an appreciable part of them) and move 
downstream.  Indeed, after crossing $z=0$ they feel
a strong quasi-perpendicular  wave field.  The 
wavenumber can be estimated as
\begin{equation}
	k_0 \simeq \frac{u_1}{u_2} k_{\rm u}
	\label{k01}
\end{equation}
where $k_{\rm u}$ is the wave number of the most unstable and
presumably strongest mode in the upstream region excited by the
escaping beam due to the cyclotron resonance
$kv_{\parallel}~+~\omega_{\rm ci}~=~0$ (we use here unnormalized
variables).  As we shall see, \( \left|v_{\parallel} \right| \) in
this resonant condition may noticeably exceed \( u_1 \).  Here, we
estimate \( k_{\rm u} \) as $k_{\rm u} \la \omega_{\rm Hi}/u_1$.
Thus, the gyroradius of particles crossing the shock downstream is
smaller than the wave length 
\begin{equation} k_0 \frac{u_1
-u_2}{\omega_{\perp}} \la \frac{u_1 -u_2}{u_2} \varepsilon < 1
\label{k02} 
\end{equation} 
and these particles must be effectively
deflected in the wave magnetic field.  Since the wave is not really
monochromatic and low amplitude turbulence is always present as well,
the motion of particles is not fully deterministic.  It should be
noted here that the background turbulence can be very well generated
by particles themselves.  One generation scenario (the so called side
band instability) has been studied for Langmuir waves in Ref.
\cite{k:d:s}, and for whistler waves in Ref. \cite{b:k:p}.  The
results can be summarized briefly as follows.  After the wave is
switched on (or, what is more appropriate for our case, the plasma has
entered the region of wave localization) and the particles have
bounced a few times in the wave field, their distribution becomes
``ergodic'' and depends only on the action $J$.  This happens due to
the fast mixing in the phase variable $\psi $.  This ergodic
distribution is, however, usually unstable with respect to the
excitation of satellites of the main wave due to the resonances
$\omega _k = n \Omega (J) $.  Here, $\omega _k$ and $k \approx k_0 $
are the frequency and the wave number of a satellite in the main wave
frame, $\Omega (J)= \partial H/\partial J $ is the frequency of
particle oscillations in the wave and $n $ is an integral number.

\begin{figure}
\TrimRight{1.5cm} \TrimLeft{0.5cm} \TrimTop{1cm} \TrimBottom{14cm}
{\bBoxedEPSF{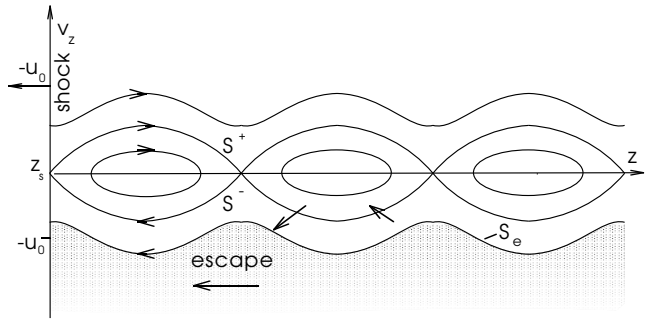}} 
\caption{Phase space of particles moving in
the wave behind a shock. The shaded region corresponds to the
adiabatically escaping particles.}  
\protect\label{fig:ps}
\end{figure}

A quasilinear theory of the backreaction of excited satellite
turbulence on the main wave and on the ergodic particle distribution has
been developed by this author \cite{m82}.  This theory shows that the
satellites change significantly both the particle distribution
and the wave.  However, they evolve relatively slowly compared with the
particle period $\Omega ^{-1}$.  In particular, the untrapped
particles can diffuse in $J $- variable and cross the separatrix $
k = 1$ becoming trapped and vice versa.  This diffusion may cover the
phase space globally in contrast to stochastic layers around
separatrices that usually develop in the case of quasi-monochromatic
perturbations \cite{zf68}. Coming back to the subject of this paper
one can expect a similar process that results in particle exchange
between the trapped and untrapped regions as shown schematically in
Fig. \ref{fig:ps}.

In a general sense the origin of the background turbulence is not important 
in our simplified model.  It is 
quite clear that there are many factors in the shock neighborhood that can 
drive such a turbulence and, hence, destroy the invariants of the particle 
motion.  Our critical assumption is, however, that the particle diffusion 
in $J$- variable associated with this turbulence is slow compared with the 
regular motion:
$	 {D(J)}/{\Omega(J)}\ll J_{\rm S}^2 $,
where $D $ is the quasilinear diffusion coefficient.                       
On the other hand we also assume that the background turbulence provides 
sufficient mixing of particles in the region $0 < z < L$, \ie 
$	D{k_0 L}/{\Omega} \gg J_{\rm S}^2 $,
so that we finally impose the following constraint on the diffusion 
coefficient
\begin{equation}
	\frac{1}{k_0 L} \ll \frac{D}{\Omega J_{\rm  S}^2} \ll 1
	\label{dif:fork}
\end{equation}
Under these circumstances the escape flux upstream becomes virtually
independent of $D$ i.e.  of the downstream background turbulence.  To
explain the point consider the phase space in Fig.  \ref{fig:ps}
again.  As seen in the wave frame the shock front $z = z_{\rm s}$ is
escaping to the left at the speed $u_{\rm 0} = u_2 - C _{\rm A} \approx
u_2$, where $C_{\rm A}$ is the phase velocity of the wave propagating
backwards in the local fluid frame as one excited by the backstreaming
particles in the upstream region and transmitted then downstream
\cite{f2}.  Therefore, the particles above the upper branch of the
separatrix $S^+, \, (k = 1) $ 
 cannot penetrate upstream.  At least some particles
below $S^- $ can in principle cross the plane $z = 0 $.  To identify
them we first consider a particle having $k(J) < 1$ below $S^- $.  Its
velocity $v_{\rm z} $ oscillates in time and depending on the
invariants $J $ and $v$ the absolute value of particle velocity can
exceed at least instantaneously the value $u_{\rm 0} $ and, therefore,
such a particle can potentially take over the shock front.  For this
to happen it must reach $z = 0$, and since the diffusion
in $J$ is slow, it must come from a far downstream region.
Therefore, it should exceed the velocity $u_{\rm 0}$ not only
instantaneously but also on average (over $\Omega ^{-1} $) to be able
to get to the shock front and then to cross it.  These particles are
shown as the dotted area in Fig.  \ref{fig:ps}.  Therefore, the area
between the separatrix $S^{-}$ and the dotted area (where $\bar v_{\rm
z} < -u_0$) should be virtually empty in the vicinity of the shock $z
= 0+$, since the shock is effectively escaping from these particles
and refilling this area by the trapped particles (above $S^-$) and
escaping (below $S_{\rm e}$) due to the diffusion across $S^-$ and
$S_{\rm e}$ is slow due to inequality (\ref{dif:fork}).  The same
arguments can be applied to the trapped particles which have $\bar
v_{\rm z} \sim \varepsilon > -u_0$ and, therefore, cannot reach the
shock.  An exception should probably be made for some particles in the
leftmost trapped region.  Their return upstream may occur when $S^-$
crosses the line \( v_{\rm z} = -u_0 \) (not shown in Fig.
\ref{fig:ps}) provided that they have a proper phase in the wave while
crossing the shock.  However, this would be rather a reflection off
the shock front than the leakage from the downstream region considered
here.  We shall not take this possibility further into account here
(see Ref.  \cite{sch98} for a reflection dominated injection
scenario).  From the above reasons we can identify the particles
escaping upstream with those below $S_{\rm e} $ in Fig.  \ref{fig:ps}.
If the amplitude of the background turbulence satisfies
condition (\ref{dif:fork}), the flux of the escaping particles can be
obtained  from the ergodic arguments regardless of any details of
their interaction with the background turbulence.  One may think of a
quasi-linear plateau that also does not depend on the form of the wave
spectrum as of a simple analog to this situation.  For the particles
on the curve $S_{\rm e} $ we have
\begin{equation}
	\bar v_{\rm z} \equiv  \frac{1}{2 \pi} \int_{0}^{2 \pi} v_{\rm z} 
	d \psi = - u_0
	\label{vz:aver}
\end{equation}
Using Eq.(\ref{eta}) and the action-angle formalism introduced in the 
preceding section one easily finds
\begin{equation}
	\bar v_{\rm z} = -\frac{\pi \sqrt{\lambda}}{k {\bf K(k)}} - \varepsilon 
	\left[\frac{1}{\lambda}- 1 + 2 \frac{{\bf E}(k) -{\bf K}(k)}{k^2 \lambda 
	{\bf K} (k)}\right]
	\label{vz:aver1}
\end{equation}
Equations (\ref{vz:aver},\ref{vz:aver1}) define a critical $k = k_*(u_0) $, 
so that particles with $k < k_* $ (dotted area in Fig. \ref{fig:ps}) 
escape upstream, whereas the rest are convected downstream.
%%********************************************************
\section{Invariant measure of adiabatically escaping particles}\label{ae}
\label{inv:mes}              
In the preceding section we have defined a boundary in particle phase space 
that divides the shocked downstream plasma into two parts, escaping 
upstream and convecting downstream.  In a four-dimensional phase space 
$({\bf v}, z)$ this boundary is a hyper-surface given by the equation (see 
Eqs. (\ref{vz:aver},\ref{vz:aver1}))
\begin{equation}
	\bar v_{\rm z}(J,\lambda) = -u_0
	\label{vz:aver2}
\end{equation}
The unperturbed motion takes place on a 2-torus labeled by the two
arbitrary chosen independent integrals of motion, which are equivalent
to two action variables.  Reducing the motion to an effectively
one-dimensional we have used the action $J$ as an action of the
one-dimensional motion, and the integral $\lambda $ accounted for the
motion in cyclic variables.  These invariants were useful for the task
of the two preceding sections.  However, our final goal is to calculate
the escaping flux starting from the Rankine-Hugoniot relations, i.e.
from the parameters of the upstream flow, which unfortunately yield
only the width of the downstream distribution in $v$, not its form,
whereas the pitch angle distribution is implied to be isotropic.  We
thus need to transform the results of the preceding section to the
variables \( v,\mu \), where \( \mu \) is the cosine of a pitch angle
in the wave frame.  A reasonable starting assumption about the
particle distribution is that far downstream where the thermalization
of the plasma (also due to the interaction with the Alfv\'en wave) is
completed, the distribution becomes isotropic in pitch angle and, for
example, a Maxwellian in $v$.

As it was argued in the preceding section the background turbulence will 
generally destroy the 2-tori and give rise to a relatively slow diffusion in 
$\lambda $ and $J$.  If we assume in addition, that this turbulence is mainly 
due to the weakly dispersive Alfv\'en waves or magnetosonic waves propagating 
almost (anti)parallel to the unperturbed magnetic field and all in the same 
direction, then this diffusion is essentially a diffusion in pitch angle.  In 
other words $v$ remains invariant and the second order Fermi acceleration is not 
important.

Our next assumption concerns mixing properties of the particle dynamics 
influenced by the background turbulence.  In particular, we assume that the 
relevant phase flux is ergodic, and that the typical length scale $L_{\bf 
\mu }$ of the pitch angle scattering satisfies the condition $1/k_0 \ll 
L_{\bf \mu } \ll L$ (see Eq.(\ref{dif:fork})).  The ergodicity implies 
that when a particle wanders in the region $z\in (0,L) $, the time spent by 
it in the ``escape'' position is proportional to the size of the ``escape'' 
region.

Thus, to calculate the fraction of the backstreaming particles that can 
cross the shock as a function of $v$ we need to calculate the fraction of 
the hyper-surface $v = {\rm const} $ downstream from the shock occupied by 
the escaping particles.  According to the preceding section this fraction 
can be represented as an invariant measure of these particles on the 
isoenergetic surface in the phase space as follows
\begin{equation}
	\nu_{\rm esc}= \frac{1}{8\pi^2}\int_{\Gamma }^{} \delta 
	\left(v^{\prime}-v\right)\, d v^{\prime}  d \mu \, d \phi \, d z. 
	\label{im}
\end{equation}
We have here introduced the spherical coordinates in the velocity space
\begin{eqnarray}
	v_{\rm z} &= &  v \mu  \nonumber \\
	v_{\rm x} & = & v \sqrt{1 - \mu^2} \cos \phi \nonumber \\
	 v_{\rm y}& = & v \sqrt{1 - \mu^2} \sin \phi
	\label{from:sph}
\end{eqnarray}
The integration region $\Gamma $ contains one wave period in $z$, i.e.  \( 
2\pi \) and is confined in other three variables by the hyper-surface $k = 
k_*(\lambda, u_0) ={\rm const}$ to be obtained from Eqs.\ 
(\ref{vz:aver},\ref{vz:aver1}).  According to the normalization used in 
Eq. (\ref{im}) the full measure $\nu (v) =1$, and Eq. (\ref{im}) thus 
yields the fraction of  particles escaping from the surface $v={\rm 
const}$.  The boundary $k_*$ of the region $\Gamma $ is not a coordinate 
surface in the variables used in Eq. (\ref{im}) over which the far 
downstream distribution is assumed to be uniform as discussed earlier in 
this section.  To evaluate the integral (\ref{im}), we therefore transform 
it to the variables already introduced in Sec.  \ref{pd} as follows: 
$(v,\mu,\phi, z) \rightarrow (\lambda, q, \eta, z) $, with
      \begin{eqnarray}
      	\lambda & = & \sqrt{v^2\left(1 - \mu^2\right) - 2 v \sqrt{1 
      	- \mu^2} \cos \alpha+1 +2 \varepsilon v \mu} \nonumber \\
      	q & = &\arctan\left(\frac{v\sqrt{1-\mu^2} \sin 
      	\alpha}{v\sqrt{1-\mu^2} \cos \alpha -1}\right)  \nonumber \\
      	\eta & = & v \mu -\varepsilon v \frac{v 
      	\left(1-\mu^2\right)-\sqrt{1-\mu^2} \cos \alpha}{1+v^2 
      	\left(1-\mu^2\right)- 2 \sqrt{1-\mu^2}\cos \alpha},
      	\label{lam:q:et}
      \end{eqnarray}  
and $\alpha = \phi + z$.	         
The absolute value of the Jacobian of this transformation can be 
conveniently expressed after some algebra through the invariants of the 
unperturbed motion $v$ and $\lambda$
\begin{equation}
	\left| \frac{\partial \left(\lambda,\eta, q\right)}{\partial 
	\left(v,\mu, \phi\right)} \right| = \frac{v^2}{\lambda} + 
	{\mathcal{O}}\left(\varepsilon^2\right)
	\label{Jac}
\end{equation}
Thus the invariant measure (\ref{im}) rewrites as
\begin{equation}
	\nu_{\rm esc} = \frac{1}{4 v^2 \pi}\int_{\Gamma}^{}\delta\left(v'
	(\lambda, \eta, q) - v\right) \lambda d \lambda \, d \eta \, d q
	\label{im1}
\end{equation}
where $v^{\prime 2} (\lambda, \eta, q) = 2 H + \left(\lambda - 1\right)^2 
$, and $H$ is defined by Eq. (\ref{Ham}).  The next transformation which 
we perform is a canonical one, also introduced in Sec.  \ref{pd}, i.e.  
$(\eta, q) \rightarrow (J, \psi) $.  Then we get:
  \begin{equation}
  	\nu_{\rm esc}=\frac{1}{2 v^2}\int_{\Gamma}^{} \delta\left(v^{\prime} 
  	-v\right) \lambda d \lambda \, d J
  	\label{im2}
  \end{equation}
Transforming this integral to the $k$- variable introduced in Sec. \ref{pd}, 
after some simple algebra we finally obtain 
  $$
      	\nu_{\rm esc} = \frac{2\sqrt{2}}{\pi v}\int_{0}^{k_*} 
      	\frac{\left(\frac{1}{2}k^2-1 +\sqrt{\frac{1}{4}k^4 v^2 -k^2 
      	+1}\right)^{{3}/{2}}}{k^3 \sqrt{\frac{1}{4}k^4 v^2 -k^2  +1}} 
      	$$
\begin{equation}
     \times	{\bf K}(k) d k
      	\label{im3}
\end{equation}
An escape of these particles is possible only for $v > 1$ and the
escape probability behaves as $ \nu_{\rm esc} \sim k_*^4 \left(v
-1\right)^{3/2}$ for $v-1 \ll 1$, where \( k_* \) is also rather small
numerically for \( v < v_{\rm th\;a} \), where $v_{\rm th\;a}$ may be
characterized as a threshold velocity of the adiabatic escape and can
be estimated from Eqs.(\ref{vz:aver},\ref{vz:aver}) as $v_{\rm
th\;a}\approx 1+\varepsilon$.  For larger \( v \), \( k_* \) rises
sharply to reach \( k_* \simeq 1 \).  For $v \rightarrow \infty$,
$\nu_{\rm esc} \approx \frac{1}{2}\left(1-v^{-1}\right)$.  The former
formula for \( \nu_{\rm esc} \) reflects the shrinkage of the phase
space of adiabatically escaping particles as \( v \rightarrow 1+\) and
does not mean that there are no other escaping particles with \( v \la
1 \) (see the next section).  The interpretation of the last formula
is straightforward: these particles are not influenced by the wave and
nearly a half of them escape.
  
To obtain the distribution of escaping particles one has to multiply 
$\nu_{\rm esc} $ by the thermal downstream distribution.  To be specific we 
assume that the latter is a Maxwellian with the downstream thermal velocity 
$v_2$.  Then, using our dimensionless velocity in the wave frame $v$ which 
is virtually the downstream velocity ($C_{\rm A} \ll u_2$) the pith-angle 
averaged distribution of escaping particles can be written 
as
\begin{equation}
	F_{\rm esc}(v) = \frac{2}{1-u_0/v} \nu_{\rm esc}(v) f_{\rm M}(v), \quad 
	v>1
 	\label{f:esc}
     \end{equation}
whereas 
     \begin{equation}
     	f_{\rm M} = \frac{n_2}{\left(2 \pi\right)^{\frac{3}{2}}v_2^3}
	\exp\left(-\frac{v^2}{2v_2^2}\right),
     	\label{maxw}
     \end{equation}     
with $v_2 = \varepsilon k_0 V_{\rm T}/\omega_{\rm ci}$, and \( V_{\rm 
T}=\sqrt{ T_2/M} \) is the downstream thermal velocity.  The factor \( 
(1-u_0/v)/2 \) accounts for the limited fraction of the phase space at 
given \( v \) in which particles escape into the upstream half-space.  
According to Eqs.\ (\ref{k01},\ref{k02}) we can estimate $v_2$ as
 \begin{equation}
 	v_2  \la \varepsilon 
	 \frac{V_{T}}{u_2} \approx \sqrt{3} \varepsilon,       
 	\label{v:2}
 \end{equation}
where the last value is valid for a strong shock with a compression 
ratio of 4.
 
 Since $\nu_{\rm esc} $ starts to grow from zero only at $v > 1$ we infer 
 that the contribution of the particles interacting with the wave 
 adiabatically is exponentially small.
%%************************
 \section{Resonant escape}\label{re}
%%************************
Let us turn now to the resonant particles, that cannot be described by the 
simple formalism developed in Sec.  \ref{pd}, since $\lambda^2 $ can be 
very small or even negative in this case.  It is convenient to use the 
variables $(v,\mu,\alpha)$ again (see Eq. (\ref{from:sph})), where $\alpha 
=\phi +z$, and a new Hamiltonian
\begin{equation}
 	{\mathcal{H}}= \sqrt{1-\mu^2} \cos \alpha +\frac{1}{2} v \mu^2 - 
 	\varepsilon \mu 
 	\label{ham:sph}
 \end{equation}
which is related to $\lambda^2$ through ${\mathcal 
H}=\left(v^2+1-\lambda^2\right)/2v$.  The exact equations then take the 
form
 \begin{eqnarray}
 	\frac{d \mu}{d t} &= &\sqrt{1- \mu^2} \sin \alpha = -\frac{\partial 
 	{\mathcal{H}}} {\partial \alpha} \nonumber\\
 	\frac{d \alpha}{d t} & = & -\frac{\mu \cos \alpha}{\sqrt{1- \mu^2}} 
+v\mu -\varepsilon= 
 	\frac{\partial {\mathcal{H}}}{\partial\mu}
 	\label{eq:sph}
 \end{eqnarray}
The particle trajectories are shown in Fig.  \ref{lamb:sqfig} as a
contourplot of the function $\lambda^2$ with fixed \( v \).  Next we
concentrate on the region of small and negative $\lambda^2$.  First,
consider the vicinity of the elliptic point of the Hamiltonian
(\ref{ham:sph}) $\alpha~=~0~\pmod {2 \pi},\, \mu~=~\mu_0$.  The
equation for $\mu_0$ reads
\begin{equation}
    	-\frac{\mu}{\sqrt{1-\mu^2}}+v \mu =\varepsilon
    	\label{mu0:eq}
    \end{equation}   
and in  the case of small $\varepsilon \ll 1$ and $\left|\mu_0\right| \ll 
1$, the last equation  reduces to a 
cubic one which yields for $\mu_0 $: 
\begin{equation}
\mu_0= \left \{ 
\begin{array}{l}
	 -2\left|\xi\right|^{\frac{1}{2}}\sinh \left[\frac{1}{3}\sinh^{-1} 
\frac{\varepsilon}{\left|\xi\right|^{\frac{3}{2}}}\right], \, 
	 \xi \le 0   \\	
	  -2\xi^{\frac{1}{2}}\cosh \left[\frac{1}{3}\cosh^{-1} 
\frac{\varepsilon}{\xi ^{\frac{3}{2}}}\right], \, 
	 0 < \xi < \varepsilon^{\frac{2}{3}}  \\
	2\xi^{\frac{1}{2}}\sin \left[\frac{1}{3}\sin^{-1} 
\frac{\varepsilon}{\xi ^{\frac{3}{2}}}-\frac{2\pi}{3}\right], \,  
	 \xi > \varepsilon^{\frac{2}{3}} 
\end{array}
\right.
	\label{mu0:sol}
\end{equation}
where $\xi=\frac{2}{3}(v-1) $.  The bottom expression  is 
strictly valid when $\xi \ll 1$, otherwise a more accurate treatment of 
Eq. (\ref{mu0:eq}) is needed.  At the same time, for larger $ \xi $ the 
main contribution to the particle escape comes from the adiabatic region 
which was already considered in Sec. \ref{inv:mes}.  Moreover, the 
downstream thermal particle distribution falls off very rapidly in $v$, and 
the behavior of the escape probability at the lower energies is generally 
more important.  We start, therefore, our consideration from the case 
\(v<1\).  A particle that moves at the critical point $\alpha~=~0,\quad 
\mu~=~\mu_0 $ can escape only when
\begin{equation}
	\mu_0~v~<~-u_0 
	\label{ineq}
\end{equation}
(see Eq. (\ref{vz:aver2})).  Since $\left|\mu_0\right|$ is always
small for $\xi~<~0$ ($\mu_0~\simeq-~\left(~2\varepsilon~\right)^{1/3}$
for \( \left|\xi\right|<\varepsilon^{2/3} \), and \( \mu_0~\simeq
2\varepsilon /3\xi \) for \( \left|\xi\right|>\varepsilon^{2/3}\)),
the last inequality can not be fulfilled for all resonant particles
with $v~<~1$ and therefore a threshold velocity $v_{\rm th}$ occurs.
As it was argued in Sec.  \ref{bt}, the dimensionless velocity $u_0$
can be estimated as
\begin{equation}
	u_0\simeq\frac{k_0u_2}{\omega_\perp}\simeq\varepsilon,
	\label{cr:vel}
\end{equation}
where $u_2$ is the bulk velocity in the downstream region (see also Eq. 
(\ref{k01}) ).  In practice $u_0$ can deviate from the value (\ref{cr:vel}) 
due to a number of reasons, \eg due to a finite propagation speed of the 
wave and/or due to the fact that \( k_0 \) depends on \( v_{\rm th} \).  
Therefore, we represent \( u_0 \) as $u_0~=~\varepsilon~\zeta$, where 
$\zeta~\simeq~1$.  Then, using Eq. (\ref{mu0:eq}) inequality (\ref{ineq}) 
rewrites as
\begin{equation}
	v > v_{\rm th} \equiv \zeta\sqrt{\left(\zeta+1\right)^{-2} 
+\varepsilon^2}\simeq 
\frac{1}{2} +\varepsilon^2
	\label{v:th}
\end{equation}
To obtain the number of particles that are trapped into nonlinear
resonance around the point $\alpha~=~0,\quad\mu=\mu_0$ and escape
upstream we can use the same arguments as in Sec. \ref{inv:mes} in
our calculation of the escape flux of the untrapped particles.  Thus,
given $v~>~v_{\rm th}$ we calculate the critical orbit around \(
\alpha~=~0,\quad \mu~=~\mu_0 \) for which the averaged velocity \(
\bar v_{\rm z} \equiv v\bar \mu=-u_0 \).  Due to the anharmonicity of
the oscillations the averaged velocity \( \bar v_{\rm z} \) that
starts from the value \( \bar v_{\rm z}=v\mu_0 \) at the bottom of
potential well, increases with the radius of the orbit, reaching
finally \( -u_0 \).  Near the threshold where \( v-v_{\rm th} \ll 1
\), the condition \( \bar v_{\rm z} = -u_0 \) is fulfilled for an
orbit which is close to the critical point of the Hamiltonian.  It is
therefore convenient to introduce a new variable \( \nu=\mu-\mu_0 \)
and expand the Hamiltonian (\ref{ham:sph}) at \( \alpha=\nu=0 \),
retaining only cubic anharmonicity in \( \left|\nu\right|\ll 1 \), and
neglecting \( \alpha^4 \) and \( \alpha^2 \nu \mu_0 \) compared with
\( \alpha^2 \ll 1 \).  Thus, the truncated Hamiltonian takes the form
\begin{equation}
	{\mathcal{H}}_1=-\frac{\alpha^2}{2}+\frac{3}{4}\left(\xi-\mu_0^2\right)
	\nu^2 -\frac{\mu_0}{2}\nu^3
	\label{H1}
\end{equation}
We introduce again the action-angle variables \( (\nu,\alpha) \mapsto 
(\psi,J) \), considering \( \alpha \) as a momentum. The transformation 
is generated by
$	S=\int_{}^{}\alpha d \nu $.
Thus, for \( \psi,J \) we obtain
\begin{equation}
	J=\oint \alpha d \nu, \quad \psi=\frac{\partial S}{\partial \alpha}
	\label{ang:act}
\end{equation}
where 
\begin{eqnarray}
	 \alpha & = & \sqrt{-\mu_0 \left(\nu^3-a_2 \nu^2 +a_0\right)}, 
	\nonumber \\
	a_2 & = & \frac{3}{2}\left(\frac{\xi}{\mu_0}-\mu_0\right)
	\label{alpha}
\end{eqnarray}
and \( a_0 \) plays the role of the ``energy'' constant of the anharmonic 
oscillator (\ref{H1}).  It is worthwhile to perform the following 
transformation
\begin{equation}
	\nu=\frac{1}{3}a_2 \left(2 z +1\right)
	\label{nu}
\end{equation}
and to rewrite \( \alpha \) in Eq. (\ref{ang:act}) as follows
\begin{equation}
	\alpha=\left(\frac{2}{3}a_2\right)^{\frac{3}{2}}\sqrt{-\mu_0 
\left(z-z_{-1}\right) \left(z_0-z\right) \left(z_1 - z\right)},
	\label{alpha1}
\end{equation}
where
\begin{equation}
	z_{\rm n}=\sin \left(\frac{1}{3}\sin^{-1} a +\frac{2 \pi n}{3}\right)
	\label{z:n}
\end{equation}
Here \( a \) again denotes the ``energy'' of the oscillator (\ref{H1}) and 
varies between -1 and 1 in the potential well.  Substituting 
Eqs.\ (\ref{nu},\ref{alpha1}) into Eq. (\ref{ang:act}), we obtain after some 
simple algebra

$$
J=3^{\frac{3}{2}}\frac{\left(\mu_0^2-\xi\right)^{\frac{5}{2}}}{5\sqrt{2}\pi 
\mu_0^2} 
$$
\begin{equation}
\times \frac{2 \left(k^{\prime 2}+k^4\right){\bf E}(k)-k^{\prime 
2}\left(1 + k^{\prime 2}\right){\bf K}(k)}{\left(k^{\prime 
2}+k^4\right)^{\frac{5}{4}}}
	\label{J1}
\end{equation}
where \( k^2=\left(z_0 -z_{-1}\right)/\left(z_1-z_{-1}\right) \) and \( 
k^{\prime 2} \equiv 1-k^2 \). As it was mentioned, there exists a critical 
\( J=J_{\rm e} \), such that the particles with \( J < J_{\rm e} \) escape, 
whereas  the particles with \( J \ge J_{\rm e} \) do not. To calculate \( 
J_{\rm e} \) we introduce the averaged \( \bar \mu \) as
\begin{equation}
	\bar \mu (J) = \frac{1}{2 \pi} \int_{0}^{2 \pi} d \psi \mu (J,\psi)
	\label{mu:bar}
\end{equation}
Then \( J_{\rm e} \) will be defined by
\begin{equation}
	\bar \mu (J_{\rm e}) v = -u_0 \equiv -\zeta \varepsilon
	\label{mu:bar:v}
\end{equation}
The last equation can be rewritten as
\begin{equation}
	v \left(\bar \nu + \mu_0\right) = -\varepsilon \zeta
	\label{esc:eq}
\end{equation}
where
$	\bar \nu = ({1}/{T}) \oint {\nu d \nu}/{\alpha (\nu)} $,
\( \alpha (\nu) \) is given by Eq. (\ref{alpha}), and 
$	T=\oint {d \nu}/{\alpha} $.
After a short calculation we obtain

$$
	\bar \nu = \frac{a_2}{\sqrt{k^{\prime 
2}+k^4}}
$$
\begin{equation}
\times \left[\frac{1}{3}\left(2-k^2+\sqrt{k^{\prime 
2}+k^4}\right)-\frac{{\bf E}(k)}{{\bf K}(k)}\right].
	\label{nu:bar1}
\end{equation}
Inserting the last expression into Eq. (\ref{esc:eq}) we first obtain \( 
k_* \) as a root of the equation
\begin{equation}
	v \left(\bar \nu (k_*)+ \mu_0\right) = -\varepsilon \zeta,
	\label{esc:eq1}
\end{equation}
and using then Eq. (\ref{J1}), we finally obtain \( J_{\rm e} = J(k_*) \). 

\begin{figure}
\TrimBottom{12.7cm}
\TrimLeft{3.7cm}
\TrimRight{3cm}
\TrimTop{1.5cm}
{\bBoxedEPSF{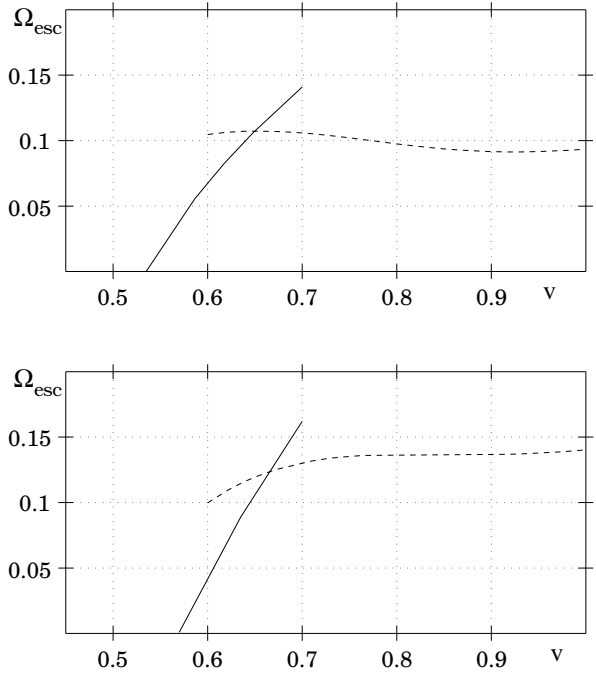}}	
	\caption{Escape probability; $\varepsilon = 0.2$ (upper panel) and 
$\varepsilon = 0.3 $ (lower panel).  Solid lines correspond to Eq. 
(\protect\ref{omega:esc2}), whereas dashed lines correspond to Eq.
 (\protect\ref{omega:esc1}) with $k_* $ from Eq. (\protect\ref{k:large}).}
	\protect\label{fig:escpr}
\end{figure}

Normalizing the full measure of the particles at \( v={\rm const} \) to unity
\begin{equation}
	\Omega_0=\frac{1}{4 \pi} \int_{-1}^{1} d \mu \int_{-\pi}^{\pi} d \alpha =1
	\label{omega)}
\end{equation}
and adopting arguments similar to those already used in Sec.
\ref{inv:mes}, we find for the invariant measure of the resonantly
escaping particles
\begin{equation}
	\Omega_{\rm esc}=\frac{1}{4 \pi} \int_{J<J_{\rm e}}^{} d \alpha d \mu
	\label{omega:esc}
\end{equation}
Transforming \( d \alpha d \mu \rightarrow d \psi d J \) we thus find
\begin{equation}
	\Omega_{\rm esc} = \frac{1}{2} J_{\rm e} = \frac{1}{2} J(k_*)
	\label{omega:esc1}
\end{equation}
Not far from the threshold velocity (\( v \ga v_{\rm th} \)), where \( 
k_* \) is small, from Eq. (\ref{esc:eq1}) we obtain
\begin{equation}
	\frac{3}{16} a_2 k^4_{*}\approx -\mu_0 -\frac{\varepsilon \zeta}{v}
	\label{k:small}
\end{equation}
Since
\[ J(k) \approx 
\frac{3^{\frac{5}{2}}}{2^{\frac{9}{2}}}\frac{\left(\mu_0^2-\xi 
\right)^{{5}/{2}}}{\mu_0^2 } k^4, \]
for \( \Omega_{\rm esc} \) we have
\begin{equation}
	\Omega_{\rm esc} \approx \frac{\sqrt{2}}{6}\left(3 \mu_0^2-2 
(v-1)\right)^{{3}/{2} } \left(1+\frac{\varepsilon \zeta}{v \mu_0}\right)
	\label{omega:esc2}
\end{equation}
which in the case \( 1-v > \varepsilon^{{2}/{3}} \) simplifies to 
 \begin{equation}
	 \Omega_{\rm esc } \approx \frac{2}{3v} (1+\zeta) 
	(1-v)^{\frac{3}{2}}\left(v-v_{\rm th}\right)
 	\label{omega:sim}
 \end{equation}
As \( v \) grows approaching unity, Eq. (\ref{k:small}) becomes invalid 
and in the case opposite to Eq. (\ref{k:small}) i.e., when \( k^{\prime} 
\ll 1 \) from Eq. (\ref{esc:eq1}), we find
\begin{eqnarray}
	k_*^2 & \approx & 1-16 \exp \left\{-3v \frac{2(v-1)-3\mu_0^2}{2(v-1) v+3 
\varepsilon \mu_0 \zeta}\right\} \label{k:large}
	 \\
	 & \approx & 1-16 \exp \left\{-3 \cdot 2^{{1}/{3}}
\varepsilon^{-{2}/{3}}/\zeta \right\}
\end{eqnarray}
Note, that the last expression is valid only for \( \left|1-v\right| < 
\varepsilon ^{4/3} \).  The escape probability then takes the form, 
Eqs.\ (\ref{J1}, \ref{omega:esc1})
\begin{equation}
	\Omega_{\rm esc}=\frac{3^{{3}/{2}}}
{5\sqrt{2}\pi}\frac{\left(\mu_0^2-\xi^2\right)^{{5}/{2}}}{\mu_0^2}
	\label{omega:esc3}
\end{equation}
The dependence \( \Omega_{\rm esc}(v) \) as calculated for small $k $ and 
for $k \simeq 1$ is shown in Fig. \ref{fig:escpr}. 
Since the downstream thermal distribution falls off very 
rapidly, the resulting escape spectrum will have a maximum in 
$v$ close to the point \( v = v_{\rm th} \approx \onehalf  \). The 
distribution of the escaping particles can be written again as 
\begin{equation}
	F_{\rm esc} = \frac{2}{1-u_0/v} \Omega_{\rm esc} f_{\rm M}(v)
	\label{esc:prob}
\end{equation}
For larger \( v \ga 1 \), when the approximation \( \left|\mu_0\right|
\ll 1 \) breaks down, formula (\ref{omega:esc3}) becomes invalid as
well, and a more accurate consideration of Eqs.\ (\ref{eq:sph}) is needed
in this case.  At the same time \( \Omega_{\rm esc}\sim \varepsilon \)
for \( v\simeq 1 \), as may be seen from Eq. (\ref{omega:esc3}), and the
escape in the region \( v-1 \ga 1 \) is dominated by particles that
interact with the wave adiabatically.  Thus, for \( v \ga 1 \) the
distribution of escaping particles can be given by Eq. (\ref{f:esc}).
It should be born in mind, however, that the maximum of the escape
distribution is rather close to \( v = \onehalf \)  for  
small \( \varepsilon \), so that formulae (\ref{omega:sim}) and
(\ref{omega:esc3}) provide virtually the kernel of the distribution of
escaping particles, whereas Eqs.\ (\ref{im3},\ref{f:esc}) describe the
tail of this distribution.
%%**************************
\section{Injection efficiency versus mass to charge ratio}\label{a:z}
%%**************************
One important aspect of any injection mechanism should be its dependence 
upon the mass to charge ratio of different species.  It is obviously so in 
the mechanism suggested in this paper.  Indeed, in the case \( \varepsilon 
\ll 1 \), the leakage upstream must be controlled by the 
parameter \( k_0 \rho_\alpha \), where \( \rho_\alpha=(V_{T\alpha 
}/\omega_\perp) (A/Z)\), is the Larmor radius of a species \( \alpha \), 
i.e.  \( V_{T\alpha} \) is a corresponding thermal velocity, \( A \) and \( 
Z \) are the mass and charge numbers, respectively.  It is clear that 
strongly magnetized particles (\( k_0 \rho_\alpha \ll 1 \)) cannot be 
injected, whereas unmagnetized particles (\( k_0 \rho_\alpha \gg 1 \) are 
injected as readily as in the case without magnetic field.  For protons this 
parameter is \( k_0 \rho_{\rm p} \equiv v_2 \ga \varepsilon \) almost by 
definition, simply due to the fact that both the wave and the thermal 
distribution downstream originate from the same upstream flow (see 
Eq. (\ref{v:2})).  To confine particles effectively the parameter \( 
\varepsilon \) must be rather small but not too small-- otherwise the 
upstream turbulence cannot be exited by a weak proton beam.  This means 
that protons are close to a watershed between the species that cannot be 
injected by this mechanism (these are apparently electrons, only) and 
particles with higher \( A/Z \) whose injection efficiency increases.

\begin{figure}
\TrimTop{2.5cm}
\TrimBottom{17cm}
\TrimLeft{3cm}
\TrimRight{3cm}
{\bBoxedEPSF{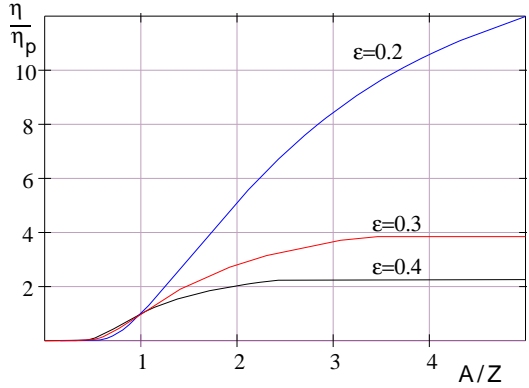}}	
	\caption{Injection efficiencies  of different species normalized to 
 proton efficiency as functions of 
 mass to charge ratio and for different wave amplitudes.} 
	\protect\label{abund}
\end{figure}

According to the preceding section the most important physical quantity 
that regulates the escape flux is the threshold velocity \( v_{\rm th} \) 
which is the same for all species, Eq. (\ref{v:th}), provided that in the 
definition of the normalized velocity \( {\bf v} \), Eq. (\ref{dimlvar}), 
one substitutes \( \omega_{\perp \alpha} = \omega_\perp \cdot (Z/A) \) 
instead of \( \omega_\perp \).  Thus, the subjects for injection are only 
the particles with \( v > v_{\rm th} \simeq 1/2 \), or \( (k_0 
v/\omega_\perp)(A/Z) > 1/2 \) for unnormalized \( v \).  The escape 
probabilities \( \nu_{\rm esc}(v) \) and \( \Omega_{\rm esc} (v) \) (Secs.  
\ref{ae},\ref{re}) as functions of dimensionless velocity are also the same for 
all sorts of particles.  The quantity that discriminates particles against 
\( A/Z \) ratio in dimensionless variables is obviously the downstream 
thermal velocity \( v_2 \), Eq. (\ref{v:2}).  For a species \( \alpha \) 
we thus have
\begin{equation}
	v_{2 \alpha}=v_2 \frac{V_{T \alpha}}{V_{T}} \frac{A}{Z}
	\label{v:alf}
\end{equation}
We may use now Eqs.\ (\ref{f:esc}) and (\ref{esc:prob}) for calculating the 
distribution of escaping particles of a sort \( \alpha \) upon substituting 
\( v_{2 \alpha} \) instead of \( v_2 \).  It is of course assumed that all 
the arguments of Sec.  \ref{bt} are valid for these particles as well.  
There exists, however, the problem of the thermal velocities \( V_{T 
\alpha} \).  It is indeed very difficult to quantify them at the current 
level of description.  The simplest assumption is that upon crossing the 
shock these particles behave more or less like the protons.  In other 
words, their excess (over \( u_2 \)) velocity, i.e.  \( u_1 -u_2 \), is 
spread around \( u_2 \) and we assume that \( V_{T \alpha} \simeq V_{\rm T} 
\).  For the purpose of simplicity and for extracting the dependence upon 
\( A/Z \) we also assume that the thermal distributions of all species \( 
\alpha \) is equivalent to that of the protons
\begin{equation}
	f_\alpha = \frac{n_{2 \alpha}}{\left(2\pi\right)^{3/2} v_{2 \alpha}^3} 
\exp \left(-\frac{v^2}{2 v_{2 \alpha}^2}\right)
	\label{f:alf}
\end{equation}
Then, we may calculate the density \cite{f3}
of injected particles
\begin{equation}
	n_{\rm inj}^\alpha = \int_{v_{\rm z} < -u_0}^{}\frac{2 d{\bf v}}{1-u_0/v} 
f_\alpha \bar \Omega_{\rm esc}
	\label{n:1}
\end{equation}
where \( \bar \Omega_{\rm esc}= \Omega_{\rm esc} \) for \( v < 1 \) and 
\(\bar \Omega_{\rm esc}= \nu_{\rm esc}  \) for  $ v > 1 $ (see 
Eqs.\ (\ref{f:esc}) and (\ref{esc:prob})). 
The last equation can be evaluated to 
\begin{equation}
n_{\rm inj}^\alpha = \sqrt{\frac{2}{\pi}} \frac{n_{2\alpha}}{v_{2\alpha} 
^3} \int_{\hat v}^{\infty} v^2 dv 
\bar \Omega_{\rm esc} (v) \exp 
\left(-\frac{v^2}{2 v_{2\alpha}^2}\right)
	\label{n:2}
\end{equation}
where $\hat v= \max\left\{v_{\rm th}, u_{0 \alpha}\right\}$ \( u_{0
\alpha} \equiv u_0 A/Z \).  According to Eqs.\
(\ref{omega:esc2},\ref{omega:esc3}) and Fig.  \ref{fig:escpr} the
function \( \bar \Omega_{\rm esc} (v) \) rises sharply from \( \bar
\Omega_{\rm esc} (v_{\rm th}) =0 \) to become approximately constant
$$ \bar \Omega_{\rm esc} \simeq \Omega_0 \equiv \frac{3\sqrt{6}}{5\pi}
\varepsilon $$ in the region \( v_{\rm th} + \sigma \varepsilon \la v
\la 1 \), where \( \sigma =3^{5/2}/10\pi \).  For \( v >1 \) the
escape flux is dominated by adiabatic particles.  To simplify the
algebra we substitute \( \bar \Omega_{\rm esc} \simeq \Omega_0 \) into
Eq.  (\ref{n:2}) and shift the lower limit \( v_{\rm th} \to v_{\rm
th}^*=v_{\rm th} +\hat\sigma\varepsilon \) where \( \hat\sigma \sim
\sigma \) and extend the integral to \( \infty \).  Starting from \(
v=1 \) we may use the high energy asymptotic result \( \nu_{\rm esc} =
(1-1/v)/2 \), multiplied by \( 1-2 \Omega_0 \) to compensate the above
extension of the contribution of resonant particles.  This simple
interpolation yields for the protons, $\eta\equiv n_{\rm
inj}^{\rm p}/n_{2 {\rm p}} $
\begin{eqnarray}
	\eta & = & 
\frac{1}{2}-\left(\frac{1}{2}-\Omega_0\right) 
\Phi\left(\frac{1}{\sqrt{2}v_{2}}\right) -\Omega_0 
\Phi\left(\frac{v_{\rm th}^*}{\sqrt{2}v_{2}}\right)  \nonumber 
\\ & + & \sqrt{\frac{2}{\pi}}\frac{\Omega_0}{v_{2}} v_{\rm 
th}^*\exp\left(-\frac{v_{\rm th}^{*2}}{2 
v_{2}^2}\right)
	\label{eta:alf}
\end{eqnarray}
where
$$\Phi(x) = \frac{2}{\sqrt{\pi}} \int_{0}^{x} e^{-t^2} dt $$
For other species one may obtain a similar formula from Eq. (\ref{n:2}). 
We illustrate it by plotting the injection 
efficiency \( \eta_\alpha  \) normalized to the proton efficiency. This 
is shown in Fig. \ref{abund} for \( 
v_{2\alpha} =v_2 A/Z \) and \( \hat\sigma =1.5 \sigma \).
%%*****************************
\section{A simplified selfconsistent model}\label{sc}
%%*****************************
So far, we have considered particle escape under a prescribed wave
spectrum downstream.  However, as we emphasized, this escape mechanism
ought to possess a very distinct selfregulation.  Indeed, there is a
strong negative feedback between the wave intensity and the density of
the escaping beam-- if the beam is weak and excites thus only weak
waves, the leakage will be increased to produce stronger waves.
Similar arguments lead to decreasing the leakage if the beam is too
strong.  Therefore, both the beam intensity and the turbulence
amplitude must rest at some definite and unique level.  What makes
this situation differ from the standard quasilinear theory of beam
relaxation in homogeneous plasmas is that ``beam relaxation'' here
means actually its return to the shock front via the cyclotron
interaction with the self-excited MHD waves (see Ref.\ \cite{mv95} for
a detailed description of this process).  Hence, a plateau does not
form in fact and the relaxation length \( l_{\rm R} \) means simply a
distance at which the majority of beam particles are turned around and
swept back to the shock.  The nonlinear wave phenomena are assumed to
be unimportant, which implies that the corresponding time scale \(
\tau_{\rm NL} > l_{\rm R}/u_1 \).  Thus, the fraction of the beam
energy that may in principle be channeled into the plasma heating
through the nonlinear wave-particle interactions is correspondingly
small and the wave energy at the shock front may be calculated from a
simple energetic balance.

First we note that the beam energy is \( mn_{\rm b}v_{\rm b}^2/2 \), where 
\( \left|v_{\rm b}\right| \ga u_1 \) is the beam velocity in the upstream 
frame and \( n_{\rm b} \) is its density.  Since it scatters back 
quasi-elastically, namely around scattering centers that move at the low 
velocity \( -C_{\rm A} \), only a \( C_{\rm A}/v_{\rm b} \ll 1 \) fraction 
of beam energy may be converted into waves.  A complete quasilinear theory 
of cyclotron beam relaxation in homogeneous plasmas has been developed in 
Ref.\ \cite{rss}.  One can show that the expression for the wave energy 
released by an unstable beam as calculated by these authors is also 
applicable for the case considered here.  Thus, for the magnetic field 
perturbation upstream we may write
\begin{equation}
	\frac{B_{\perp u}^2}{8\pi} \simeq \frac{\Lambda}{4} 
n_{\rm b}m_{\rm p} C_{\rm A}v_{\rm b}
	\label{am:1}
\end{equation}
We have merely introduced an additional factor \( \Lambda \sim \Delta 
v_{\parallel}/v_{\rm b} <1 \) where \( \Delta v_\parallel \) is the beam 
width in parallel velocity.  This factor appears because the beam 
relaxation occurs under the constraint of conservation of the (zero) 
particle flux \( \int_{}^{}v_{\parallel }f_{\rm b} d v_{\parallel }\) on a 
given diffusion line \( v_\perp ^2 +(v_{\parallel }+C_{\rm A})^2 = const \) 
in velocity space rather than under the conservation of the phase density 
\( \int_{}^{} f_{\rm b}d v_{\parallel }\) along this line.  The reason for 
such a factor \( \Lambda \) may be understood from the observation that 
particles starting to escape at velocities \( \sim -\Delta v_{\parallel } 
\), while being turned around can hardly acquire positive velocities that 
are appreciably larger than \( +\Delta v_{\parallel } \) before returning 
to the shock.  Again, because the above mentioned particle flux must be 
zero.  Since the downstream field \( B_\perp =rB_{\perp u} \), where \( r = 
u_1/u_2\) is the shock compression ratio we may rewrite Eq. (\ref{am:1}) 
as follows
 \begin{equation}
 	\frac{1}{\varepsilon^2} \simeq \frac{1}{2}r^3 \Lambda M_{\rm A} 
\eta
 	\label{am:2}
 \end{equation}
where \( M_{\rm A}\simeq v_{\rm b}/C_{\rm A} \) and \( \eta =n_{\rm b}/n_2 \) is 
given by Eq.  (\ref{eta:alf}).  Since \( \eta(\varepsilon ) \) is a 
monotonically increasing function, Eq.  (\ref{am:2}) determines a unique value 
of \( \varepsilon \) and thus a unique injection rate \( \eta \).

Consider first the case \( M_{\rm A} \gg 1 \). To the leading 
approximation in \( \varepsilon \ll 1 \) and substituting \( v_2 = 
\sqrt{3}\varepsilon \),  \( r=4 \) into Eq. (\ref{eta:alf}) we obtain
\begin{equation}
\eta = \frac{6}{5\pi^{3/2}}v_{\rm th}^* \exp\left(-\frac{v_{\rm 
th}^{*2}}{6\varepsilon^2}\right)
	\label{eta2}
\end{equation}
Eq. (\ref{am:2}) then rewrites
 \begin{equation}
 	\frac{1}{\varepsilon^2}=C M_{\rm A}\exp\left(-\frac{v_{\rm 
th}^{*2}}{6\varepsilon^2}\right)
 	\label{am:3}
 \end{equation}
where 
$$C=\frac{192}{5\pi^{3/2}}v_{\rm th}^*\Lambda  $$
For \( CM_{\rm A} \gg 1  \), and assuming $C \sim 1 $ we thus obtain
\begin{equation}
	\varepsilon^2 \simeq {v_{\rm th}^{*2}}/{6L_{\rm A}}
	\label{am:4}
\end{equation}
where $$
L_{\rm A} = \ln\left(M_{\rm 
A}\frac{v_{\rm th}^{*2}}{6\ln M_{\rm A}}\right).$$
For the injection rate \( \eta \) we have
\begin{equation}
	\eta = \frac{3 L_{\rm A}}{16\Lambda M_{\rm A}v_{\rm th}^{*2}}
	\label{eta:fin}
\end{equation}
One may put here \( v_{\rm th }^* \simeq 1/2 \). It is seen that the 
injection rate formally  vanishes with \( 1/M_{\rm A} \) as \( \eta 
\sim M_{\rm A}^{-1}\ln M_{\rm A} \). However, already for \( \varepsilon 
\la 1/4 \), \( B_{\perp u} \sim B_{\rm z} \) and the estimate of the wave 
amplitude might need some correction. Clearly, the above \( \eta \) 
scaling is not applicable in the limit \( B_{\rm z} \to 0 \) in which \( 
\eta \)  vanishes. For such a weak magnetic fields different mechanisms 
of beam relaxation must be considered. 

\begin{figure}
\TrimTop{6cm}
\TrimBottom{12cm}
\TrimLeft{3cm}
\TrimRight{3cm}
{\bBoxedEPSF{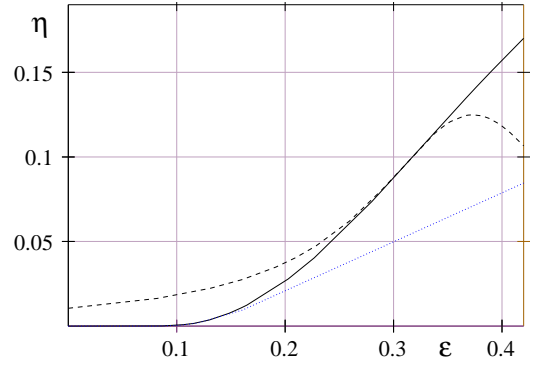}}	
	\caption{The actual injection rate calculated from Eq. (\protect\ref{eta:alf}) 
(solid curve); the approximations given by Eq. (\protect\ref{eta2}) (dotted curve) 
and Eq. (\protect\ref{eta:ap}) (dashed curve). } 
	\protect\label{selfc}
\end{figure}

\begin{figure}
\TrimTop{6cm}
\TrimBottom{12cm}
\TrimLeft{3cm}
\TrimRight{3cm}
{\bBoxedEPSF{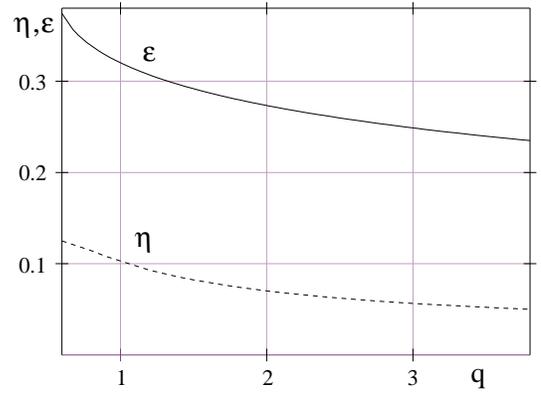}}	
	\caption{The parameter $\varepsilon $ (solid curve)  and injection 
rate $\eta $ (dashed curve) as functions of $ q=32\Lambda M_{\rm A} c_1 $.} 
	\protect\label{fin}
\end{figure}

For moderate values of \( \varepsilon < 1 \), Eq. (\ref{am:2}) can in
principle be easily solved numerically with \( \eta (\varepsilon) \)
given by Eq. (\ref{eta:alf}).  We know already that the above formulae
are applicable for rather small values of \( \varepsilon \la 0.2 \).
On the other hand \( \varepsilon \) cannot be too large in any case, in
fact it cannot be larger than about 0.4 to satisfy Eq. (\ref{am:2}).
Thus, what is actually needed is a reasonable but simple approximation
of \( \eta(\varepsilon ) \) in Eq. (\ref{am:2}) in the interval \( 0.2 \la
\varepsilon \la 0.4 \) to resolve Eq. (\ref{am:2}) for \( \varepsilon
\).  It is convenient to use the following approximation
\begin{equation}
	\eta=\frac{c_1}{1-c_2 \varepsilon +c_3 \varepsilon^2}
	\label{eta:ap}
\end{equation}
which is shown in Fig. \ref{selfc} for \( c_1=0.0105;\,
c_2=4.91;\,c_3=6.58 \).  Denoting  \( q=32\Lambda M_{\rm A} c_1 \) we
obtain the following solution for \( \varepsilon \) as a function of \(
q \)
\begin{equation}
	\varepsilon =
\frac{c_2}{2(c_3-q)}\left[1-\sqrt{1-\frac{4}{c_2^2}(c_3-q)}\right]
	\label{eps:ap}
\end{equation}
It is shown in Fig. \ref{fin} together with \( \eta(q) \). One 
sees that the injection rate depends rather slowly on \( M_{\rm A} \) as in 
the case of higher \( M_{\rm A} \) (smaller \( \varepsilon \)) considered 
earlier.
%%**************************************
\section{The leakage process in a nutshell}\label{ns}
%%**************************************
The mechanism of ion leakage considered in the previous sections
unfortunately requires more calculations than seems to be appropriate
to its physical simplicity.  At a phenomenological level this
mechanism is almost as simple as the escape from an oblique shock
where for a particle to catch the shock it must move at the speed \(
u_2/\cos \Theta_{nB_2}\simeq u_2/\varepsilon \) along the field line.
Here \( \Theta_{nB_2} \) is the angle between the downstream magnetic
field and the shock normal).  Such an escape has been extensively
studied in Ref.\ \cite{edm:k:eichl}.  Even if the shock is
quasiparallel but the magnetic field is locally oblique to the shock
normal most of the time, the same kinematic escape condition holds for
magnetized particles.  As we have seen in Sec. \ref{a:z}, most of the
protons must be magnetized and in the case of the turbulence dominated
by a circularly polarized Alfv\'en wave the lowest energy particles
that can escape have a velocity \( v=v_{\rm th} \simeq 1/2 \) ($v_{\rm
th} \simeq u_2/2\varepsilon $ in unnormalized variables).  According
to the phase plane shown in Fig.  \ref{lamb:sqfig} these are the
particles that move towards the shock being close to the point where
\( \alpha =\phi+z=0;\, \mu=\mu_0 <0 \).  Therefore, they fall into the
cyclotron resonance with the wave (\( \phi+z=0, \) \(\phi=- z=-\mu_0
vt \) but they spiral as electrons in the unperturbed field \( B_{\rm
z} \) trying to follow the magnetic field line and to minimize thus
the Lorenz force, the only force in our model that can prevent their
escape.  The inclination of their orbit to the \( z- \) axis is,
however, about a half of that of the magnetic field: \(v_\perp/v_{\rm
z} \equiv \sqrt{v_{\rm x}^2+v_{\rm y}^2}/v_{\rm z} \simeq
(1-v)/\varepsilon \simeq 1/2\varepsilon\) whereas \( B_\perp/B_{\rm z}
\equiv 1/\varepsilon \).

As we have seen in Sec. \ref{a:z}, in the case of very small values of \( 
\varepsilon \) these particles make the bulk of the leakage.  Since they 
are concentrated in a relatively small region of the downstream phase 
space, this allows us to forecast their energy and angular distribution, 
just as they appear upstream.  First, their energy per mass in the 
downstream frame must be somewhat above \( E = v_{\rm th}^2/2 \), where \( 
v_{\rm th }\simeq 1/2 \) in the dimensionless variables.  Furthermore, 
since \( v_\perp/v_{\rm z} \simeq 1/2\varepsilon \gg 1 \) the energy of 
leaking particles is predominantly in the perpendicular motion.  In the 
unnormalized variables we thus have
\begin{equation}
	{\mathcal{E}} \simeq {\mathcal{E}}_\perp 
\ga \frac{m}{2}\frac{\omega_\perp^2}{k_0^2}v_{\rm
	th}^2 > \frac{mu_1^2}{2}\frac{v_{\rm th}^2}{\varepsilon^2 r^2}
\label{en:est}
\end{equation}
whereas the parallel energy may be estimated as \(
{\mathcal{E}}_{\parallel } \simeq 4 \varepsilon^2 {\mathcal{E}}_\perp
\). It should be noted that the pitch angle scattering upstream may
change this relation to a certain extent. On the other hand it is in a
reasonable agreement with the results of the   strong shock 
simulation \cite{quest,be95}.
%%*********************************** 
\section{Injection efficiency. Comparison with hybrid simulations}\label{comp}
\begin{figure}
	\TrimRight{1.5cm} \TrimLeft{0.5cm} \TrimTop{0cm}
	\TrimBottom{14cm} 
	{\bBoxedEPSF{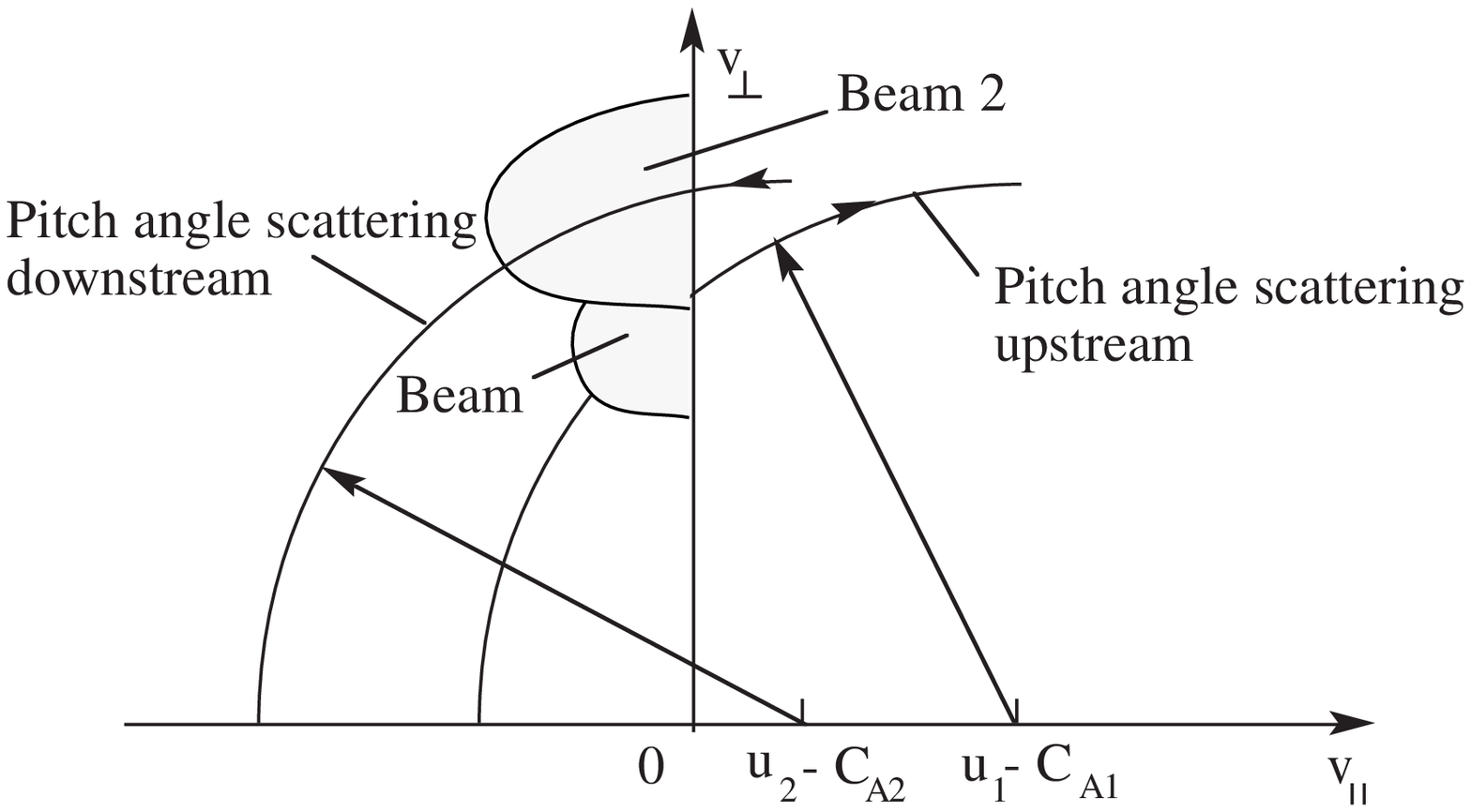}} 
	\caption{Velocity
	space of particles just in front of the shock in the shock
	frame of reference.  The ``beam'' particles schematically
	represent the leakage from the thermal distribution
	downstream.  After being scattered in pitch angle first
	upstream and then downstream, as shown by the arrows, these
	particles appear in front of the shock again but at higher
	energies and overlapping with beam particles.  They are shown
	as ``beam 2''.}  \protect\label{b:ps}
\end{figure}

%%*********************************** 
As we emphasized in the Introduction the calculation of the flux of
leaking particles alone does not solve the problem of injection.  The
main result of injection theory should be the high energy asymptotics
of a spectrum that emerges in a steady state when the leaking
particles repeatedly cross the shock and achieve energies sufficient
for describing them by the means of the standard theory of diffusive
shock acceleration (see \eg ref.  \cite{dru83}).  The mathematical
formalism of injection theory has been developed in Ref.  \cite{mv95}.
Now we may apply it to the distribution of leaking particles
(thermostat distribution) calculated in the present paper.  We also
have to bear in mind that particles that cross the shock more than
once (higher generation of injected particles, beam 2, etc., see
Fig.\ref{b:ps}) are still subject to the filtering on their way back
upstream due to the interaction with the downstream trailing wave.  As
we have seen this interaction weakens with the energy and the
thermostat becomes transparent to particles with $v \gg \omega_\perp
/k_0$ (unmagnetized particles).  With this in mind, the whole
algorithm may be outlined as follows.

Suppose that some fraction of the downstream plasma leaks upstream to
form at $z=0$ the one-sided distribution $F({\bf v}), v_\parallel =
v_{\rm z}' < 0$ (see Fig.  \ref{b:ps}).  Here ${\bf v}'$ is the
velocity in the shock frame and we keep our notation ${\bf v}$ for the
wave frame in the downstream medium (almost the downstream frame).
Due to pitch angle scattering in the upstream medium these particles
turn around and eventually cross the shock in the downstream direction
forming the distribution $F^+({\bf v}),\ v_{\rm z}' > 0$ which can be
written as $F^+=L_1 F$, again at \( z=0 \).  The linear operator
$L_1$, the upstream propagator, can be obtained from the solution of
the kinetic equation \cite{mv95}.  According to the thermostat model
in use (Secs.  \ref{intr1},\ref{bt}) these particles penetrate further
downstream through the thermostat mixing up with the hot downstream
plasma.  At the same time they are pitch-angle scattered on the
background turbulence so that some of them acquire negative velocities
and move back to the shock.  We denote their distribution within the
thermostat by \( F^- \).  For \( F^- \) we thus have
\begin{equation}
	F^-=L_2 L_1 F +f_{\rm M}
	\label{maseq1}
\end{equation}
Here $L_2$ is the downstream propagator, and $f_{\rm M}$ is the
distribution function of the downstream thermal plasma that emerges
upon the first crossing of the shock interface (without higher
generations).  In Secs.  \ref{ae},\ref{re} we assumed for simplicity
that $f_{\rm M}$ is a Maxwellian distribution so that $L_2 f_{\rm
M}\approx f_{\rm M}$ because it is isotropic in the wave frame. Now
the calculation of the injection spectrum that appears just upstream
of the shock is nothing more than the calculation of the spectrum of
leaking particles already made in Secs. \ref{ae},\ref{re} with $f_{\rm
M}$ in Eq. (\ref{esc:prob}) replaced by $F^-$ from Eq. (\ref{maseq1}).
Thus, the distribution of injected particles for $F$ takes the form
\begin{equation}
	F=\tau L_2 L_1 F+\tau f_{\rm M}.
	\label{maseq2}
\end{equation}
Here the function $\tau(v)$ is given by (see Eq.(\ref{esc:prob}))
\begin{equation}
	\tau(v)=\frac{2}{1-u_0/v} \nu_{\rm esc}
	\label{tau}
\end{equation}
and may be interpreted as a thermostat transparency coefficient.
According to Sec.  \ref{ae} $\tau \to 1$ as $v \to \infty$.  If there
were no scattering upstream ($L_1=0)$, Eq.(\ref{maseq2}) would be
equivalent to Eq.  (\ref{esc:prob}) that yields the distribution of
escaping particles given the thermal distribution downstream, $f_{\rm
M}$.  In general, Eq.  (\ref{maseq2}) is an integral equation for $F$.
The kernel $L_2 L_1$ has been calculated in \cite{mv95} where also the
solutions of this equation have been studied for $\tau \equiv 1$ and
various functions $f_{\rm M}$ which should have mimicked the effect of
thermostat filtering ($\tau < 1$).

\begin{figure}
\TrimBottom{17cm}
 	\TrimTop{3cm}
 	\TrimRight{0cm}
 	\TrimLeft{6cm}
% 	\hSlide{2cm}
%  	\vSlide{0.5cm}
{\bBoxedEPSF{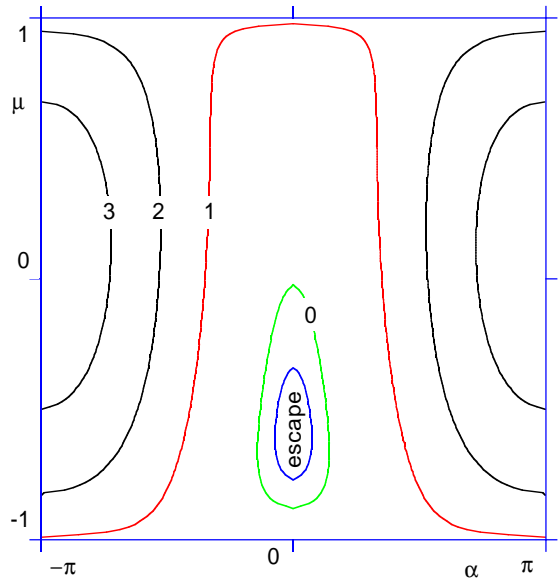}}	
\caption{ Contourplot of $\lambda^2$ for $v=0.9$ and $\varepsilon = 0.2$.}
	\protect\label{lamb:sqfig}
\end{figure}

Expression (\ref{tau}) implies only an adiabatic leakage which is
appropriate for particles with $v > 1$, Sec.  \ref{ae}.  If the
resonant particles leak as well, the escape probability $\Omega_{\rm
esc}$ from Eq.  (\ref{esc:prob}) should be added to $\nu_{\rm esc}$ in
Eq.  (\ref{tau}) and it will dominate in the region $v \la 1$.
However, unlike the leakage of adiabatically interacting particles,
the leakage of the resonant particles is very sensitive to the wave
polarization.  This may be understood from inspection of Fig.
\ref{lamb:sqfig}  drawn for an A-wave.  The MS-wave case
can be obtained by flipping the phase portrait in Fig.~\ref{lamb:sqfig} 
since the MS polarization corresponds to $\epsilon <
0$ and the Hamiltonian (\ref{ham:sph}) is invariant to the
transformation $\epsilon\to -\epsilon, \, \mu \to -\mu$.  Thus, the
candidates for the resonant leakage from an MS-wave would be the
particles marked by 2 and 3 and them alike, \ie those circulating
around a fixed pont at $\alpha =-\pi \pmod {2\pi}, \, \mu > 0$ in Fig.
\ref{lamb:sqfig}.  However, they have relatively low values of
$\bar{\mu}$ and rough estimates show that they cannot escape.  At the
same time, according to our discussion of the leakage from the
thermostat in Sec.  \ref{bt} such particles can potentially escape
from the region immediately behind the shock front provided that the
wave field is sufficiently perturbed.  This point should be born
in mind when we compare our results with hybrid simulations below.

Most of the hybrid simulations are essentially time dependent, since
the shock runs through a finite spatial domain.  Eq.  (\ref{maseq2})
implies a steady state and for this rather preliminary comparison we
select only a hybrid simulation \cite{be95} where the simulation box
was anchored on the shock front and a quasi-stationary spectrum was
developed.  A more thorough comparison with other numerical results
will be done elsewhere.  As we have seen, the most important
parameters that determine the distribution of leaking particles are
the amplitude, the wave number and the polarization of the trailing
wave.  The injection spectrum is then formed depending primarily on
the shock compression and to some extent on the spectrum of the
background turbulence upstream and downstream that enters the
propagators $L_1$ and $L_2$ in Eq.  (\ref{maseq2}) \cite{mv95,mv97}.
For the purpose of comparison we assume the MS polarization as
observed in \cite{be95} (see, however ref.\cite{f1}) and therefore
discard the contribution of resonantly escaping particles.

All other quantities needed for calculation of the injection spectrum
can be obtained from the above results given $M_{\rm A}$ and $M_{\rm S}$ and
from RH conditions.  However, shocks that form in simulations do not
follow the latter exactly, due to the losses through the boundaries of
simulation box and other reasons discussed earlier.  Therefore, it is
appropriate to take some critical parameters directly from
simulations.  For example, the total compression ratio obtained in
\cite{be95} is close to $4.2$ exceeding the limiting value of 4.  At
the same time the shock is noticeably modified, so that the local
compression ratio is substantially smaller.  We make our comparison
taking $r=4.0$ which also corresponds to a strong shock.  The
downstream temperature slightly deviates from the RH prescriptions as
well and we take it from the simulations ($T=2\cdot 10^6K$) in order
to ensure coincidence of the thermal parts of the spectra.  The
amplitude parameter $\varepsilon$ for the shock of $M_{\rm A}=5.25$ in
\cite{be95} may be calculated using Eq.  (\ref{am:2}) or
(\ref{eps:ap}).  We estimate $\Lambda = \Delta v_{\parallel}/v_{\rm b}
\approx 1$ which is appropriate for a strong shock, and obtain
$\varepsilon \approx 0.3$.  This $\varepsilon$ is in reasonable
agreement with the simulation results.  It should be reminded however,
that Eq.  (\ref{eps:ap}) was derived for the resonant leakage and a
somewhat different although very similar equation should have been
used in the case of adiabatic leakage which would have given
a somewhat higher $\varepsilon $.  However,  some of the
resonant particles probably  leak in the simulations and we adopt Eq.
(\ref{eps:ap}) for our estimate of $\varepsilon$.  Furthermore, in the
selfconsistent determination of $\varepsilon$ in Sec.  \ref{sc} only
the ``first generation'' particles are taken into account which
clearly leads to an overestimated $\varepsilon$ in Eq.  (\ref{am:2}).

We calculate the wave number $k_0$ from the condition of frequency
conservation across the shock transition, $k_0=k_{\rm u}r (M_{\rm
A}-1)/(M_{\rm A}-\sqrt{r})$, where $k_{\rm u} $ is an upstream wave
number that we in turn obtain from the resonant condition $k_{\rm u}
v_{\rm b} \approx \omega_{\rm ci}$ that we wrote here to the leading
order in $1/M_{\rm A}$.  In general, the diapason of unstable wave
numbers may be quite broad since the escaping beam is broad in
$v_{\parallel}$.  An accurate calculation of the most unstable $k$ is
a difficult problem since the beam distribution depends on this $k $
as well and we restrict ourselves to a simple estimate based on the
mean beam velocity.  Namely, the escaping beam occupies in velocity
space at least an interval $-v_{\rm th\;a}\omega_{\perp}/k_0+u_2 \la
v_{\parallel} < 0$ where $v_{\parallel}$ is a dimensional particle
velocity in the shock frame and $v_{\rm th\;a}$ is a demensionless
threshold velocity for an adiabatic escape (see text below Eq. (\ref{tau})),
$v_{\rm th\;a} \approx 1+\varepsilon$ roughly independent of the wave
polarization.  From this considerations, we obtain $\omega_\perp/k_0
u_1 \approx 1.1$.

We compare our analytic calculations (see Appendix for more detailes)
with the hybrid simulations \cite{be95} in Fig.  \ref{comp:fig}.  The
downstream Maxwellian is the same in both cases and it is drawn with
the thin line.  The squares are from simulation whereas the heavy line
shows the result of integration of Eq.  (\ref{maseq2},\ref{spectr}).  The slope of
the energy spectrum $\propto E^{-\sigma}$, with $\sigma =
3/(r-1)\approx 1$ that must form at energies sufficiently higher than
the thermal energy according to the standard theory of diffusive shock
acceleration is also shown by the dotted-dashed line drawn at arbitrary
height.  Finally, the dashed line shows the solution of Eq.
(\ref{maseq2}) with $\tau \equiv 1$, \ie for a completely transparent
thermostat \cite{f4}.

The first conclusion that may be drawn from Fig.  \ref{comp:fig} is
that the effect of particle filtering by the thermostat is very strong
and reduces the injection rate by one order of magnitude compared to
the case of ``free'' injection, \ie without the strong wave particle
interaction downstream.  Furthermore, the agreement with the
simulation spectrum is very good, although the latter does not exhibit
a correct high energy asymptotics, most probably due to the
losses of high energy particles.  It is reasonable to assume that if
there where no such losses a correct spectral slope (with probably
somewhat higher amplitude) would be achieved in the simulation at an
energy where the slope has now a minimum 
(intersection point with  the analytical spectrum).  This means that the two methods
would produce similar injection rates if the latter is
understood as an amplitude of the high energy asymptotics $\propto
E^{-\sigma}$.  The lack of low energy particles in the analytical
spectrum should be attributed to the absence of resonantly leaking (or
reflected) particles that are probably still present in the simulations.  This
might also slightly underestimate the injection rate as it is perhaps
the case in the simulations because of the losses.  At the same time
we feel that whatever physical ingredients (like the resonantly
leaking particles) are added to the above calculation scheme it will
not change the injection rate significantly.  This is due to the
strong selfregulation of the leakage (injection) process.  The shock
seems to leak always at a critical rate which is just enough to
confine the downstream plasma through wave generation as it was
discussed earlier. This fundamental aspect of particle injection at
quasi-parallel collisionless shocks has been foreseen by previous authors, 
\eg \cite{park61,quest,dmv89}.
%%***********************************
\section{Other possible approaches, existing and prospective}\label{other}
%%*********************************** 
Existing theories of shock dissipation and shock acceleration have not
included the injection of the shocked plasma into the foreshock region
selfconsistently.  An
important insight provide hybrid simulations, but being substantially
limited in space, time and particle energy, they miss
the backreaction of accelerated particles on the shock structure and, therefore, 
on injection and shock dissipation.
Monte Carlo simulations (\eg \cite{el85}) include the backreaction but they
completely ignore the feedback from the turbulence excited by injected
particles themselves which may reduce the injection rate by an order
of magnitude without major changes in the flow structure.

Although the collisionless shock phenomenon is very complicated, the
necessary information about the source of leaking ions can be
inferred from the ordinary jump conditions.  The latter show that at
least behind a strong shock the plasma distribution is so broad that a
very large fraction of it can in principle escape upstream.  This will
certainly smear out the subshock as a distinct structure, unless this
thermal return is choked by fast unstable coupling with the incoming
flow.  Thus, the problem is not the source of the particles to be
injected upstream, but rather the opposite, i.e.  how to confine most
of them on the downstream side of the shock and, of course, how to
calculate the distribution of the rest which is leaking.

One may attempt to do this in several ways.  Firstly, one can invoke
the processes occurring at the very shock interface and immediately
behind the shock, before the thermalization of the plasma flow is
completed.  One obvious candidate is the electrostatic barrier
appearing \eg at the subshock in the high Mach number hybrid
simulations \cite{quest} (some further discussion can be found \eg in
a review paper \cite{scu95}).  This is critically important in the
quasiperpendicular shock mechanism \cite{ler:etal}.  Quest
\cite{quest} noticed, however, that its impact on inflowing ions in
strong quasiparallel shocks is very modest due to the remarkably
perfect compensation of the electrostatic force with the Lorenz force.
On the other hand, this conclusion may not be true for the
backstreaming ions.

The next possibility consists in the already mentioned subshock reduction by a 
pressure gradient built up by the intense return beam in front of the shock.  
This is the mechanism at work in Monte Carlo simulations \cite{el85}.  In fact, 
it is the only mechanism of selfregulation of the injection process at the 
subshock level in Monte Carlo models since they operate under a prescribed 
scattering law.  It is, however, not to be confused with the process of a large 
scale shock modification by diffusively accelerated high energy particles which 
can also reduce the subshock.  Although these two processes of shock 
modification are physically very similar, the latter operates over a much larger 
spatial scale and depends, besides the injection rate, on factors that have 
nothing to do with the subshock physics, like losses at highest energies 
\cite{m97a}.  Note that the selfregulation mechanism suggested in this paper 
works very efficiently regardless of (or along with) the flow modification in 
the precursor and the above-mentioned feedback from the high-energy particles.

The simplified model considered here gives explicit formulae for the
distribution of backstreaming particles upon the wave amplitude
(through the parameter \( \varepsilon \)).  The latter has, in turn,
been calculated by considering the transformation of beam energy into
the wave energy.  This upstream wave, driven by the unstable beam, may
also be a subject to one of the known saturation mechanisms,
especially in the case when a compressional component of the wave
field is added.  Besides the quasilinear beam relaxation considered
here, these may be wave steepening \cite{ken:malk}, other processes of
nonlinear wave transformation, or beam trapping (see \eg Ref.\
\cite{sg69}).  Nevertheless, like in the case considered here, the
wave amplitude should be calculated as a function of the beam
intensity, providing the injection efficiency with no
parameterization.
%%%%%%%%%%%%%%%%%%%
\section{Limitations}\label{lim}
%%%%%%%%%%%%%%%%%%%%
We have assumed the unperturbed particle motion to be determined by a
monochromatic wave which is an extreme idealization in the shock
environment.  At the same time it is often the case in wave-particle
interaction, that the form of the wave field is not important for
particle trapping-- only the depth of a `potential well' is important.
The worrying situation here might be created by resonantly escaping
particles (Sec.  \ref{re}), that seem to require a `fine tuning' in
wave-particle interaction.  On the other hand, the background
turbulence that was assumed to be sufficiently strong
(Eq. (\ref{dif:fork})), certainly diminishes the role of this fine
tuning by destroying the integrals of regular dynamics.  Put another
way, a particle escapes not because it stays in exact resonance with
the wave for a long time which would hardly be possible for any
realistic wave field at a shock, but because it appears at the right
place in phase space while being close to the shock front.  Otherwise
its motion may be quite irregular.  The key element of our treatment
that allowed us to calculate the escape flux under a restricted
knowledge of chaotic particle dynamics was, of course, the ergodicity
assumption.

The injection scheme presented here will require certain modification when 
applied to the case of finite $\theta_{\rm nB}$.  Scholer \etal demonstrated by 
means of hybrid simulations \cite{sch98}, that the contribution of particles 
staying sufficiently long at the shock front is increasingly important in this 
case.  Within our scheme, these particles can be formally identified with the 
particles that are in nonlinear resonance with the trailing wave, have the 
averaged velocity $\bar{v}_{\rm z}\approx -u_0$ (marginal escape) and are close 
to the shock front.  Upon interaction with it they gain energy, although the 
mechanism wherby it hapens is yet to be studied.

Generally, the tight link between the escape flux and the wave
amplitude emphasized in this paper is the essence of the
selfregulation of the shock dissipation process given \eg by
Eq. (\ref{am:1}).  However, depending on the age and size of the shock
this may be not the only way the shock regulates its own energy
dissipation and particle acceleration.  In nonlinearly accelerating
shocks the subshock strength may be significantly reduced.  Also, the
deceleration of the flow in front of the shock by high energy
particles drives the subshock Mach number to lower values which may
have an important impact on both the injection and the overall flow
structure near the shock.

%%%%%%%%%%%%%%%%%%%%%%%%%%%%%%%
\section{Conclusions and discussion}\label{concl}
%%%%%%%%%%%%%%%%%%%%%%%%%%%%%%%
We have demonstrated that the large amplitude wave-train can
efficiently filter the warm downstream plasma in its leakage upstream,
scattering back typically no more than 5 percent of the downstream
protons in the case of a strong shock and a left-hand polarized (Alfven) wave. 
MS-type of polarization results in noticeably better confinement of the hot 
downstream plasma and  weaker leakage (injection). This means that the 
MS-turbulence is more suitable for maintaining a distinct quasi-parallel shock 
structure than the A-turbulence. The spectrum of high energy particles
accelerated out of the backstreaming beam is calculated with
the help of injection theory \cite{mv95}. The
resulting spectra are: (i) in reasonable agreement with the broad
dynamical range hybrid simulations to date \cite{be95} (ii) they
evolve into a standard power-law at higher energies (iii) their
intensity may easily exceed the threshold of the nonlinear
acceleration regime (see below).

It is needless to say that a reliable calculation of injection rate, that 
take account of all essential interrelations between physical processes like the 
leakage/reflection, wave generation, particle trapping and shock modification by 
energized particles, could dramatically improve our understanding of how strong 
shocks accelerate particles.  Recent analytic solutions \cite{m97a,m97b} for 
nonlinearly accelerating shocks (i.e.  shocks whose structure may be almost 
entirely determined by accelerated particles) show that the dependence of the 
acceleration efficiency upon the injection rate has a critical character 
allowing for extremely different solutions at quite close or even the same injection 
rates.  Therefore, the studies of the energetic particle (cosmic ray) production 
\cite{bl:eichl,axf94,dru95,vlk97} in such shocks or, in other words, of how the shock 
distributes its energy between thermal and nonthermal components of the shocked 
plasma should perhaps be focused on the subshock where particles are 
injected into the acceleration process.  The necessary subshock parameters 
should, however, be determined selfconsistently from kinetic nonlinear 
calculations of the shock structure like those mentioned above.  
\section*{acknowledgments} 
I would like to thank Heinz V\"olk for many fruitful 
discussions of various aspects of diffusive shock acceleration theory.  Exchange 
of ideas with Don Ellison was very useful as well.  I am also indebted to him 
for furnishing detailed results of numerical simulations, both Monte Carlo 
and hybrid.  This work was done within the SFB 328 of the DFG. 
%%%%%%%%%%%%%%%%%%%%%
\appendix
\section*{}
In an expanded form Eq.(\ref{maseq2})  can be written as (see \cite{mv95} for further 
details) 
\begin{eqnarray}
	F(v) & = & \frac{2\pi^2(\kappa \kappa_1)^{2/3}}{3^{1/3}\Gamma ^2(2/3)} 
	\tau \frac{\zeta_+^2}{\zeta_-^2} \int_0^{\eta_0} \eta d \eta 
{\mathrm Ai } 
	 \left (\kappa_1^{1/3}\frac{\zeta_+}{\zeta_-} \eta \right ) 
	\nonumber \\
	 &  & {\mathrm Ai }  \left (\kappa^{1/3}  \frac{\zeta_+}{\psi}\eta  \right ) 
\cdot F (v_1) + \tau f_{\rm M}(v)
	 	\label{spectr}
\end{eqnarray} 
 Here the following notations have been used 
 $$\zeta_\pm = v \pm u_2, $$ 
 $$ v_1 =\sqrt{v^2 -2\Delta u \left (\zeta_+ \eta + \psi \hat
\eta\right )}, \quad \Delta u=u_1-u_2,$$
 $$\psi = \sqrt {u_1^2 +\zeta_+ \zeta_- -2\Delta u \zeta_+ \eta } - u_1,$$ 
 $$\eta_0 =  \min (1,\zeta_-/2\Delta u ), $$
  $\Gamma $ is the Gamma function, Ai denotes the Airy 
 function, and $\hat \eta \simeq 2/3$.

The coefficient $\kappa_1 $ is a certain functional of the spectral
density of the background turbulence downstream, that ensures pitch
angle scattering and have been discussed already in Sec.  \ref{bt}.
We take $\kappa_1$ from \cite{mv95}, Eq.(67), using slightly different
notations
\begin{equation}
	\kappa_1 = \frac{a^3}{\Lambda_{\rm
	d}}\left\{2\cosh\left[\frac{1}{3}\cosh^{-1}(a^{-3}-1)
	\right]-1\right\}^3 \label{kap1}
\end{equation}
where 
$$a = {\Gamma(1/3) \over 3^{4/3} \Gamma(2/3) \Lambda_{\rm d}^{2/3}}$$ 
and 
$$\Lambda_{\rm d} = \frac{1}{\zeta_-}\int_0^{\zeta_-} d \zeta (1 -
\zeta^2/\zeta_{-}^{2})^2 / D(-\zeta)$$ 
Here $D$ is the diffusion
coefficient in velocity space normalized to its value at $\zeta=0$ as
a function of the resonant ($\zeta \approx u_2 +\omega_{\rm ci}/k$) parallel
velocity of the ions in the shock frame.

The coefficient $\kappa$ has a similar meaning as $\kappa_1$ but it is related to the 
particle transport in the upstream medium.
\begin{eqnarray}
	\kappa & = &\frac{2}{3\Lambda}protect\left\{3-\gamma +2\sqrt{\gamma (6-\gamma)} \times\right.
	\nonumber \\
	 &  & \left. 
	 \sinh\left[\frac{1}{3}\sinh^{-1}\left(\frac{-\gamma^2+9\gamma-27/2}{\sqrt{\gamma} 
	 (6-\gamma)^{3/2}}\right)\right]\right\}
	\label{kap}
\end{eqnarray}
where
$$\gamma=\frac{4\pi^3\sqrt{3}}{27\Gamma^6(2/3)\Lambda^2}$$
and 
$$\Lambda= \frac{1}{\zeta'_-}\int_0^{\zeta'_-} d \zeta (1 -
\zeta^2/\zeta^{\prime 2}_-)^2 / D(-\zeta)$$ 
Here $\zeta'_-=V-u_1$ is an
upstream analog of $\zeta_-$ calculated for the upstream absolute
value of particle velocity $V$.  Note that in \cite{mv95} the
expression for $\kappa_1$ (Appendix C) has been given erroneously only
for the case $\gamma > 6$ (even for $\gamma \gg 1$) and the formula
for $\gamma $ contained a misprint.  At the same time the case $\gamma
< 6 $ has been actually considered in numerical examples, however,
with the correct numerical values of $\kappa_1$ and $\gamma$.
Generally, there still exists some arbitrariness in choosing the
parameters $\kappa$ and $\kappa_1$ since the details of the spectra of
background turbulence are not determined in injection theory
\cite{mv95}.  Nevertheless, the resulting particle spectrum may be calculated
because it is rather insensitive to parameters $\kappa$ and
$\kappa_1$, although they influence slightly the slope of the spectrum
at high energies.  In example given in Sec.  \ref{comp} we put
$\kappa=1.3$, $\kappa_1=1.15$, ignoring their possible dependence on
particle energy.  Note, that the case $\kappa\approx\kappa_1\approx 1$
corresponds to a simple assumption $D=const$ which is reasonable for
low energy part of the spectrum, $\zeta_- \ll u_2$. In this case a
slightly softer spectrum is produced at high energies for sufficiently
high downstream temperature. This was shown in \cite{mv95} where such
values of $\kappa$ and $\kappa_1$ have been employed.

\begin{figure}
\TrimRight{1.5cm} 
\TrimLeft{0.5cm} 
\TrimTop{0cm} 
\TrimBottom{0cm}
{\bBoxedEPSF{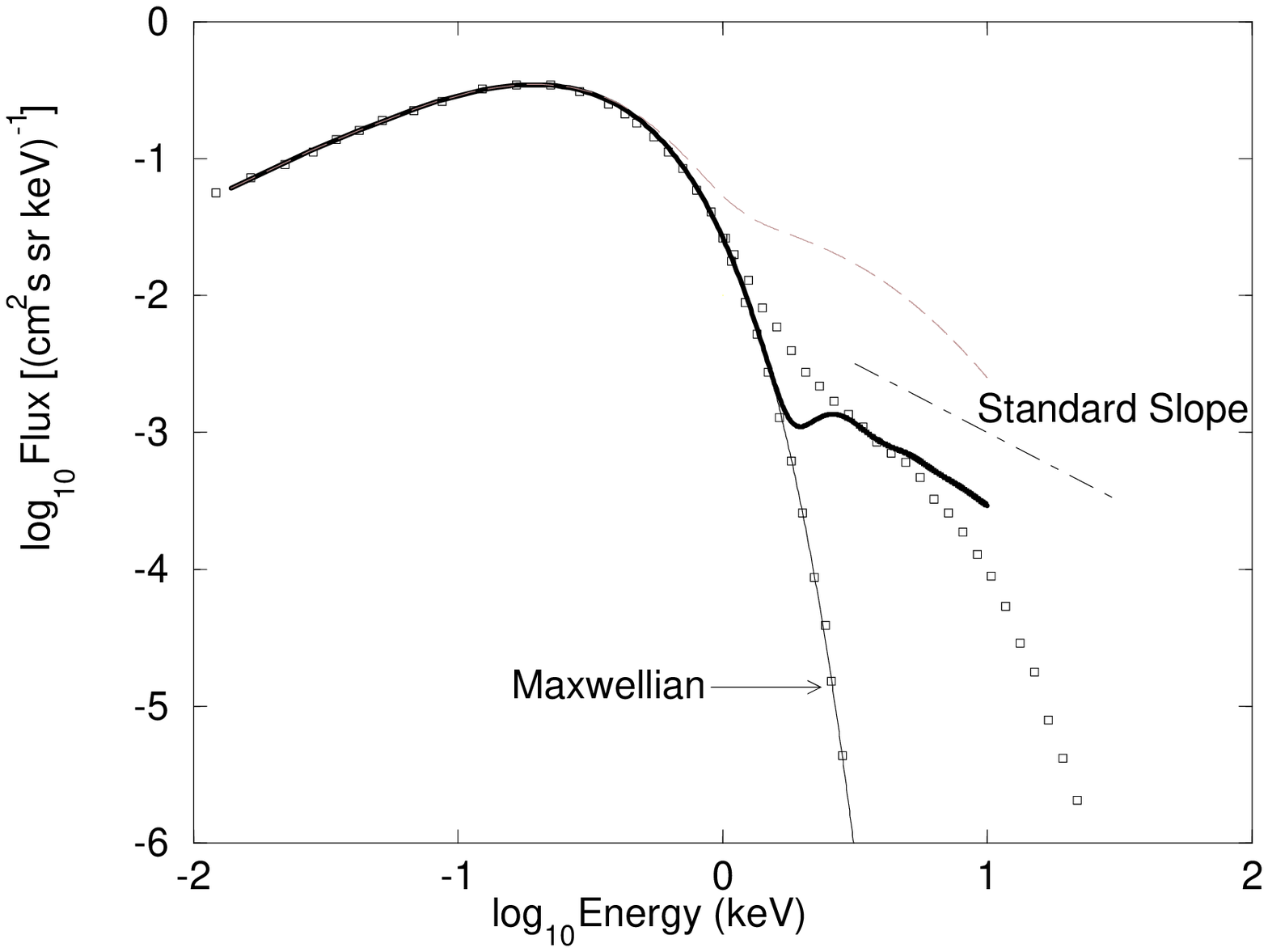}} 
\caption{Particle spectra behind the shock,
pitch angle averaged in the shock frame. The squares are from the hybrid
simulations \protect\cite{be95}. Thin line is a correspondent Maxwellian fit
which is taken as a source term $f_{\rm M}$ in Eq.(\protect\ref{maseq2}). The
solution of this equation is shown with the heavy line. The dashed line
shows the solution of the same equation for $\tau \equiv 1$. 
The dotted-dashed line indicates the slope of the spectrum appropriate for
high energy particles and shock compression of 4.}
\protect\label{comp:fig}
\end{figure}

The solution $F(v) $ in Eq.(\ref{spectr}) is in fact a one-sided
($v_\parallel < 0$ in the shock frame) isotropic (in the downstream
frame) part of the distribution function calculated at $z=0$. To
perform the matching with the high energy standard (fully isotropic)
power-law spectrum, a far downstream spectrum must be obtained since
only the latter is isotropic in the downstream frame also at lower
energies.  The necessary formulae are given in \cite{mv95}. For the
purpose of comparison with the simulation spectra given in
\cite{be95}, this far downstream spectrum is pitch-angle averaged in
the shock frame and drawn in Fig. \ref{comp:fig}.

%fig1
\input boxedeps.tex     
\SetOzTeXEPSFSpecial
\SetDefaultEPSFScale{470}
\HideDisplacementBoxes 

%%fig2
\SetOzTeXEPSFSpecial
\SetDefaultEPSFScale{430}
\HideDisplacementBoxes 

%%fig3
\SetDefaultEPSFScale{1000}

%%fig4
\SetDefaultEPSFScale{700}

%%fig5
\SetDefaultEPSFScale{600}

%%fig6

%%fig7  

%fig8
\SetOzTeXEPSFSpecial
\SetDefaultEPSFScale{470}
\HideDisplacementBoxes 
          

\begin{thebibliography}{}

	\bibitem{park61}
	E. N. Parker, J. Nucl. Energy {\bf C2}, 146 (1961)
 
	\bibitem{sag66} 
	Sagdeev, R.Z. 1966, Cooperative phenomena and shock waves in 
	collisionless plasmas,  Rev. Plasma Phys., 4, Consult. Bur., New 
	York
	
	\bibitem{ks67} 
	C. F. Kennel and  R. Z. Sagdeev, \jgr\  {\bf 72}, 3303 (1967)
	
	\bibitem{cor70}
	F. V. Coroniti, \jpp {\bf 4}, 265 (1970)

	\bibitem{lee82} 
	M. A. Lee, \jgr\ {\bf 87}, 5063 (1982)

	\bibitem{keh85} 
	C. F.  Kennel, J. P.  Edmiston, and T.  Hada, in {\it A Tutorial Review}, 
	Geophys.  Monogr.  Ser. {\bf 34}, edited by R.  G.  Stone and B.  T.  
	Tsurutani, (AGU, Washington, D.C., 1985), p.~1.


	\bibitem{pap85}
	K. Papadopoulos, in {\it A Tutorial Review}, 
	Geophys.  Monogr.  Ser. {\bf 34}, edited by R.  G.  Stone and B.  T.  
	Tsurutani, (AGU, Washington, D. C., 1985), p. 59.

	\bibitem{sch} 
	M. Scholer, in {\it Colisionless Shocks in the 
	Heliosphere: Review of Current Research}, Geophys. Monogr. Ser. {\bf 35}, 
	edited by  R. G. Stone and B. T. Tsurutani, AGU (Washington, D.C., 1985)

	\bibitem{quest} 
	K. B. Quest, \jgr\ {\bf 93}, 9649 (1988)
	
	
	\bibitem{als77}
	W. I. Axford,  E. Leer,  and G. Skadron,  Proc. 15th ICRC, Plovdiv 
	 {\bf 11}, 132 (1977)
	 
 	\bibitem{dv81}
	L. O'C. Drury,  and H. J.  V\"olk, \apj\ {\bf 248}, 344 (1981)

	\bibitem{je} 
	F.C. Jones and  D.C. Ellison, Space Sci. Rev. {\bf 58}, 259 (1991)
	
	\bibitem{mv95}
	M. A. Malkov and H. J. V\"olk, \aap\ {\bf 300}, 605 (1995)

	\bibitem{bell78} 
	A. R. Bell, \mnras\ {\bf 182}, 147; 443 (1978)
	
	\bibitem{dru83} 
	L. O'C. Drury, Rep. Prog.	Phys., {\bf	46}, 973 (1983)

	\bibitem{ll:hd} L.D.	Landau, E.M.  Lifshitz,  
	{\it Fluid Mechanics} (Pergamon, 1987)
	
	\bibitem{kenetal84}
	C. F. Kennel \etal, \jgr\ {\bf 89}, 5436  (1984) 

	\bibitem{rus:far}
	C. T. Russel and M. H. Farris, \asr {\bf  15}, 285 (1995)

	\bibitem{rss}
	J.  Rowlands, V.  D. Shapiro, and V. I. Shevchenko, Sov. Phys. JETP 
	{\bf 23}, 651 (1966)

	\bibitem{lk90}
	L. H. Lyu and J. R.	Kan, \grl\ 17, 1041	(1990)

	\bibitem{k-vo93}
	D. Kraus-Varban	and	N. Omidi, \grl\	{\bf 20}, 1007 (1993)
 
	\bibitem{sch93} 
	M. Scholer, \jgr {\bf 98}, 47 (1993)
	
	\bibitem{be95} 
	 L. Bennett and D. C. Ellison,  \jgr\ {\bf 100}, A3, 3439 (1995)
	
	\bibitem{skj} 
	M. Scholer, H. Kucharek, and V. Jayanti, \jgr {\bf 102}, 9821 (1997)

	\bibitem{bgk} 
	I.  B.  Bernstein, J.  M.  Green, and M.  D.  Kruskal, \pr\  {\bf 
	108}, 507 1957

	\bibitem{m97c} 
	M. A. Malkov, to be submitted to \jgr

	
	

	\bibitem{f1}
	

It should be noted that the magnetosonic (MS) component is also
observed and may even prevail in 1D and 2D simulations \cite{skj}.
The latter, however, are computationally very expensive and therefore
rather incomplete.  We have chosen an Alfv\'en (A) type of
polarization for a number of reasons.  First, in a real 3D situation
the MS waves experience a strong Cherenkov damping for even slightly
oblique propagation and thus they should occupy a significantly
smaller volume in $k$-space than that of the Alfv\'en waves for
similar generation and spectral transformation processes.  Second, the
criterion of modulational stability is satisfied for A-waves in the
case \( C_{\rm A} < C_{\rm S} \) unlike for MS- waves (\( C_{\rm A} >
C_{\rm S} \)), see \eg M. Longtin, and B. U.\"O Sonnerup, \jgr {\bf
91}, 6816 (1986).  We will concentrate exclusively on the case $M_{\rm
A} \gg 1$ \ie the plasma downstream is a high $\beta$ plasma ($C_{\rm
A} \ll C_{\rm S}$) and the magnetosonic wave must be modulationally
unstable.  However the above arguments and stability criteria may be
applied to the complicated strongly turbulent shock environment only
with care since they have been obtained for homogeneous plasma and in
the framework of an essentially perturbative approach.  However the
main and in fact very simple reason for choosing the A-type
polarization is that the calculation of the distribution of leaking
particles is practically very similar for A and MS waves.  Moreover,
the A-wave allows some additional group of particles to leak upstream
(see Sec.  \ref{re}) and in this sense the MS case is a subset of the
A-case.  We will use this in Sec.  \ref{comp} while comparing our
results with hybrid simulations.

	\bibitem{prig62}
	I. Prigogine, Nonequilibrium Statistical Mechanics (Wiley, New-York, 1962)


	\bibitem{sg69} 
	 R. Z. Sagdeev and  A. A. Galeev 
	{\it Nonlinear Plasma Theory }  (Reading, Mass., 1969)

	\bibitem{lut:sud} 
	R. F. Lutomirski and R. N. Sudan \pr\ {\bf 147}, 156 (1966) 

	\bibitem{on}
	T. O' Neil, Phys. Fluids {\bf 8}, 2255 (1965)

	\bibitem{k:d:s} 
	 W. L. Kruer, J. M. Dawson, and R. N. Sudan, \prl\ {\bf 23}, 838 (1969);
	V.  D.  Shapiro and V.  I.  Shevchenko Zh.  Eksp.  Teor.  Fiz. {\bf 57}, 
	2066 (1969) (Sov.  Phys.  JETP {\bf 30}, 1121, 1970)

	\bibitem{b:k:p} 
	N. I. Budko, V. I.  Karpman, and O. A. Pokhotelov,  
	Cosmic Electrodynamics {\bf 3}, 165 (1972)

	\bibitem{m82}
	M. A. Malkov, Sov. J. Plasma Phys. {\bf 8}, 495 (1982)


	\bibitem{zf68}
	G. M. Zaslavsky  and N. N.  Filonenko,
	 Sov. Phys. JETP {\bf 27}, 851 (1968)     

	\bibitem{f2}
%%ffffffffffffffffffffffffffffffffffffff
A reflected component of the wave field that is also known to appear upon 
the shock crossing has a substantially lower amplitude than that of the 
transmitted wave, see 
J. F. McKenzie and  K. O. Westphal, Planet.Space Sci. {\bf 17}, 1029 (1969);
A.  Achterberg, R. D.  Blandford, \mnras\ {\bf 
218}, 551 (1986).  We do not take it into account in our analysis of the 
regular particle motion but instead, we attribute it to the background 
turbulence.
%%ffffffffffffffffffffffffffffffffffffff
\bibitem{sch98} M. Scholer, H. Kucharek, and K. J. Trattner, \asr {\bf 21},
533 (1998)
%%fffffffffffff
\bibitem{f3}
This is, in fact, the density of particles that cross the shock 
from the downstream side for the first time, the `first generation' of 
injected particles \cite{mv95}. Subsequently, they become  subject to 
first order Fermi acceleration to form a population with a higher 
density. This process results, generally speaking, in a further 
transformation of abundances at higher energies. We also note that an 
injection rate that can be attributed to the 
beam density (\ref{n:1}) is not to be confused with the injection rate 
that is normally 
used in studies of the diffusion-convection equation where only the 
particles with momentum \( p > p_{\rm 
inj} \gg m_\alpha(u_1-u_2)\) are regarded  as 
`injected'. Here \( p_{\rm inj} \gg m_\alpha(u_1-u_2)\) is an `injection' 
momentum, a somewhat 
artificial boundary between the thermal and nonthermal plasma and only 
the latter is assumed to obey the diffusion-convection equation (see 
Refs.\ \cite{mv95,m97a} for  relevant discussions). There are also certain 
peculiarities in numerical schemes of injection, see \eg 
H. Kang and T. W. Jones,  \apj {\bf 447}, 994 (1995).
%%fffffffffffff

	\bibitem{edm:k:eichl} 
	J.P.  Edmiston, C.F.  Kennel, and D.  Eichler, \grl\ 
	{\bf 19}, 531 (1982)

	\bibitem{mv97}
	M. A. Malkov and H. J. V\"olk,  \asr {\bf 21}, 551  (1998)

	\bibitem{f4}
	In this case the injection rate is so high that
	the actual downstream temperature becomes substantially larger
	than it was assumed to be in the source term $f_{\rm M}$ in Eq. 
	(\ref{maseq2},\ref{spectr}) and the whole solution contradicts to
	the R-H conditions. To avoid this inconsistency the temperature in the 
	source term  $f_{\rm M}$ has been parametrized (lowered) in Ref.
	\cite{mv95}.

	\bibitem{dmv89} 
	 L.O'C. Drury,	W.J. Markiewicz,  H.J. V\"olk, \aap, {\bf 225} 179, (1989)

	\bibitem{el85}
	D. C. Ellison,  \jgr\ {\bf 90}, 29 (1985)

	\bibitem{scu95}
	J. D. Scudder, Adv. Space Res. {\bf 15}, 181 (1995) 

	\bibitem{ler:etal} 
	 M. M. Leroy \etal \jgr\ {\bf  87}, 5081 (1982)

	\bibitem{m97a} 
	M. A. Malkov, \apj\ {\bf 485}, 638 (1997)

	\bibitem{ken:malk}
	C. F. Kennel \etal Sov. Phys. JETP Lett. {\bf 48 }, 79 (1988);
	M.  A. Malkov \etal Phys.~ Fluids, {\bf B 3},  1407 (1991);
	M. V. Medvedev and P. H. Diamond, Phys. Plasmas, 3, 863 (1996)
 
	\bibitem{m97b} 
	M. A. Malkov, \apj\ {\bf 491}, 584 (1997)
	
	\bibitem{bl:eichl} 
	R.  D. Blandford and D. Eichler,   \physrep\  {\bf 154}, 1 (1987)

	\bibitem{axf94}
	W. I. Axford, \apjs\ {\bf 90}, 937 (1994)

	\bibitem{dru95}
	L. O'C. Drury, Adv. Space Res. {\bf 15}, 481 (1995)
	
	\bibitem{vlk97}
	H. J. V\"olk, to appear in the Proceedings of the Kruger National Park
	Workshop on TeV Gamma-Ray Astronomy (1997)






            

\end{thebibliography}
\end{document}